\documentclass[aps,prb,reprint,showpacs,longbibliography]{revtex4-1} 
\usepackage{xcolor}

\usepackage{natbib}
\usepackage{latexsym,amsmath,amssymb,amsfonts,amsthm,enumerate,url}

\usepackage{hyperref}
\hypersetup{
  colorlinks=true,
  urlcolor=blue}

\usepackage{graphicx}
\usepackage{dcolumn}
\usepackage{bm}

\newcommand{\red}[1]{{#1}}
\newcommand{\nn}{\nonumber}

\renewcommand{\(}{\left (}
\renewcommand{\)}{\right )}
\renewcommand{\[}{\left [}
\renewcommand{\]}{\right ]}
\newcommand{\la}{\langle}
\newcommand{\ra}{\rangle}
\newcommand{\dg}{\dagger}
\renewcommand{\o}{\omega}
\newcommand\rmd{{\rm d}}
\newcommand\rme{{\rm e}}

\newcommand{\beqs}{\begin{equation*}}
\newcommand{\eeqs}{\end{equation*}}
\newcommand{\be}{\begin{equation}}
\newcommand{\ee}{\end{equation}}
\newcommand{\bea}{\begin{eqnarray}}
\newcommand{\eea}{\end{eqnarray}}

\newcommand{\reff}[1]{(\ref{#1})}

\newcommand{\eq}[2][]{
\begin{equation}
#2 \label{#1}
\end{equation}}

\newcommand{\eqa}[1]{
\begin{align}
#1
\end{align}}

\begin{document}

\title{Statistics of the work done by splitting a one-dimensional \red{quasi}-condensate}

\author{Spyros~Sotiriadis$^{1,2}$,  Andrea Gambassi$^{2,3}$ and Alessandro Silva$^{2,4}$ }
\affiliation{$^1$Dipartimento di Fisica dell'Universit\`a di Pisa and INFN, Largo B. Pontecorvo 3, 56127 Pisa, Italy}
\affiliation{$^2$SISSA -- International School for Advanced Studies, via Bonomea 265, 34136 Trieste, Italy}
\affiliation{$^3$INFN -- Istituto Nazionale di Fisica Nucleare, sezione di Trieste}
\affiliation{$^4$Abdus Salam ICTP, Strada Costiera 11, 34151 Trieste, Italy}


\date{\today}

\begin{abstract}

Motivated by experiments on splitting one-dimensional \red{quasi}-condensates, we study the statistics of the work done by a  quantum quench in a bosonic system. We discuss the general features of the probability distribution of the work and focus on its behaviour at the lowest energy threshold, which develops an edge singularity. A formal connection between this probability distribution and the critical Casimir effect in thin classical films shows that certain features of the edge singularity are universal as the post-quench gap tends to zero. Our results are quantitatively illustrated by an exact calculation for non-interacting bosonic systems. The effects of finite system size, dimensionality, and non-zero initial temperature are discussed in detail.
\end{abstract}

\maketitle

\section{Introduction}

The study of non-equilibrium dynamics in thermally isolated quantum system is one of the most active areas of research in many-body quantum physics, mainly driven by ground-breaking experiments with ultra-cold atoms\cite{Dziarmaga2010,Polkovnikov2011,Lamacraft2012}. Early experiments demonstrated the occurrence of collapse and revival phenomena in systems of strongly correlated bosons, suggesting a high degree of quantum coherence despite the presence of strong many-body interactions\cite{greiner2002b}. On the other hand, the peculiar form of the interactions and in particular their integrability has been conjectured to be the reason for the observed lack of thermalization in the non-equilibrium dynamics of quasi one-dimensional bosonic gases\cite{kinoshita}. In turn, the closeness to integrability may cause the phenomenon of pre-thermalization\cite{Kollar2011,Berges2004,Gring2012,Kitagawa2011} recently studied experimentally by looking at the dynamics of the statistics of the interference contrast  of split
one-dimensional \red{quasi}-condensates on an atom chip\cite{Gring2012}. 
Atom chips, more than standard optical lattices, are particularly suited for determining probability
distributions since they allow the separate observation
of the dynamics of each individual split \red{quasi}-condensate.

Generically, the non-equilibrium dynamics is expected to depend crucially on the specific features of the system considered as well as on the way it is taken out of equilibrium\cite{Polkovnikov2011}. Among the many types of protocols studied, the simplest one is perhaps the so-called quantum quench\cite{calabrese_06}, consisting in a rapid change in the parameters of the Hamiltonian of the quantum system. Not only can this protocol be rather easily implemented in experiments, but  its theoretical description also reveals deep and interesting connections with boundary statistical field theory and confined classical systems\cite{calabrese_07,Gambassi2011,Gambassi2011a,GS12}. The effects of non-equilibrium protocols on many-body systems can be characterized in a variety of ways, e.g., by studying the time dependence of correlation functions of local operators\cite{Polkovnikov2011}. However, from a fundamental point of view, a quantum quench is analogous to
a thermodynamic transformation and as such it should be 
characterized by three macroscopic variables: the work $W$ done on the system\cite{Silva2008,GS12,Heyl2012a}, the entropy $S$ produced\cite{Polkovnikov2011,Dorner2012}, 
and the heat $Q$ possibly exchanged\cite{Polkovnikov2008}. Focusing on the work $W$ done on the system upon performing the quench, a common feature of both classical and quantum non-equilibrium processes
is that $W$ fluctuates among different realizations of the same protocol\cite{Campisi2011,Talkner2007,Jarzynski1997}. 
Accordingly, its description requires 
the introduction of a probability distribution $P(W)$. Studying $P(W)$ offers several advantages since this distribution turns out to be connected to other quantities that have been extensively studied theoretically in a variety of different physical contexts, such as the critical Casimir effect\cite{Gambassi2011} and the Loschmidt echo\cite{Silva2008}. These connections highlight emergent universal properties of systems close to criticality\cite{Silva2008,Gambassi2011} even for more general protocols\cite{Bunin2011,Smacchia2012,WSLL}. On the other hand, work is a fundamental observable and should be experimentally accessible either directly by spectroscopic methods\cite{Heyl2012a, Silva2008,Smacchia2012} or indirectly by integrating the information extracted through the cloud expansion data of cold atom experiments.

Our aim here is to discuss the qualitative features of $P(W)$  expected for global quantum quenches of extended systems on the basis of the connections between quench dynamics and boundary statistical mechanics\cite{calabrese_06,Gambassi2011,Gambassi2011a,GS12}. As we show below, the function $P(W)$ consists of a superposition of peaks which correspond to the excited states and
which gradually overlap to  form a  distribution as the system size increases. While the distribution $P(W)$ for ``large" 
values of the work $W$ depends on the microscopic details of the system, a universal behaviour might emerge\cite{Gambassi2011} 
in the form of a threshold edge singularity when $W$ is comparable with the energy of the lowest excitations. This edge singularity, which has an exponentially small spectral weight in the thermodynamic limit, is a relevant feature in systems of finite size such as those currently investigated in experiments. Our general considerations are exemplified by detailed calculations of the 
statistics 
of the work in non-interacting bosonic systems in $d$ spatial dimensions subject
to an abrupt change of the energy of the lowest excitation -- which we will refer to as ``mass" --  from $m_0$ to $m$.
This simple model is of experimental relevance in various situations, e.g., for split one-dimensional \red{quasi}-condensates 
in which the degrees of freedom associated with the relative phase of the two split components
behave at low energies as non-interacting bosons. We
also investigate the effect of a non-zero initial temperature\cite{SCC09} on the properties of the statistics of the work.

The rest of the presentation is organized as follows: In section~\ref{sec:1} we introduce general concepts and discuss a possible experimental realization of the model we study. In section~\ref{sec:2} we present the calculation of the statistics of the work for a
quantum quench and derive some of its quantitative features. In section~\ref{sec:3} we explore in detail the connections with the critical Casimir effect.
In section~\ref{sec:4} we discuss the case of a quench originating from a thermal initial state. We summarize and discuss our results in section~\ref{sec:5}.

\section{Experimental motivation, model and qualitative results}\label{sec:1}

An interesting experimental realization of the quantum quench protocol, suitable for 
studying statistical fluctuations and probability distributions, is provided by the splitting of one-dimensional (1$d$) quasi-condensates, an experiment that has been performed extensively in the context of condensate interferometry over the last few years\cite{Gring2012}. In these experiments, a system of ultra cold atoms trapped in a cigar-shaped potential and forming a 1$d$ quasi-condensate is split abruptly into two 1$d$ quasi-condensates, parallel to each other, by rapidly raising a potential barrier between them. 
The relative phase fluctuations of the two split \red{quasi}-condensates are the relevant low-energy degrees of freedom and their 
dynamics is described, under the circumstances discussed below, by a quadratic 
field theory. Indeed, while a 1$d$ system of bosons with short-range interactions may be described by the Lieb-Liniger model\cite{Cazalilla2011}, 
its low-energy properties are effectively captured by the hydrodynamic or Luttinger 
liquid approach. The coupling of two such 1$d$ quasi-condensates due to tunneling across the potential barrier which separates them can in turn be modelled by a Josephson-like coupling. As a consequence, the Hamiltonian $H$ describing their relative phase $\phi(x)=\phi_1(x)-\phi_2(x)$ 
turns out to be\cite{Gritsev2007,Gritsev_07a} that of the sine-Gordon model
\eq[hamSG]{H= \frac12 \int {\rm d}x \, \[\Pi^2 + (\partial_x \phi)^2 - 4 \Delta \cos (\beta \phi ) \] ,}
where $\Pi(x)$ is the momentum conjugate to the bosonic field $\phi(x)$. The first two terms correspond to the Luttinger model while the last one $-4\Delta \cos (\beta \phi)$ represents the Josephson coupling. The parameter $\Delta$ measures the strength of the coupling between the \red{quasi}-condensates. Indicating by $M$ the mass of the bosons, by $g$ their interaction strength (related to the scattering length) and by $\rho$ their  
number density in 3$d$,
for small values of the dimensionless parameter $\gamma=Mg/(\hbar^2 \rho) < 1$, $\Delta$ is proportional to the hopping $t$ between the two \red{quasi}-condensates and to the density: $\Delta\simeq t \rho$. The parameter $\beta$ in  Eq.~(\ref{hamSG}) is equal to $\sqrt{2\pi/K}$ where $K$ is the Luttinger liquid parameter which is related solely to $\gamma$ and varies from $+\infty$ for $\gamma\ll 1$ (weakly interacting bosons) to $1$ for $\gamma\to\infty$ (hard-core bosons or Tonks-Girardeau limit)\cite{Cazalilla2011}. More specifically %, in the small $\gamma$ limit, 
$K$ increases like $K\simeq \pi/\sqrt{\gamma} ( 1-\sqrt{\gamma}/(2\pi) )^{-
1/2}$ upon decreasing $\gamma$ towards zero. In the limit of small interactions, i.e., $\beta \ll 1 $ (or, equivalently, large $K$ or small $\gamma$), the system can be described by the quadratic approximation
\eq[hamG]{H= \frac12 \int {\rm d}x \, \[\Pi^2 + (\partial_x \phi)^2 + m^2 \; \phi^2 \] + \text{const.}}
characterized by a ``mass" \red{$m =\beta \sqrt{2\Delta}$} for the quantum field $\phi$.
Therefore we expect that the low-energy statistics of the work due to a partial, or total split of the two \red{quasi}-condensates can be described via this non-interacting bosonic model, which we mostly focus on below.
In $d$ spatial dimensions, the Hamiltonian in Eq.~\reff{hamG} is diagonalizable in independent momentum modes
\eq[ham1]{H(m) = \int_k{\(\frac{1}{2} \pi_{k}\pi_{-k} + \frac{1}{2} \omega^{2}_{k}(m) \phi_{k}\phi_{-k}\)},}
where $\[\phi_k,\pi_{k'}\]=(2 \pi)^d i\delta_{k,k'}$. By $\int_k$ we denote $\int_{\rm BZ}{\rm d}^dk/(2\pi)^d$, i.e., the integral over the first Brillouin zone in the case of a lattice model, which becomes the whole $d$-dimensional space in the continuum limit. In the following we actually consider only this limit, although our discussion can be readily extended to the case on the lattice. Henceforth we assume a relativistic dispersion relation $\o_k(m)=\sqrt{k^2+m^2}$ and focus on quenches of the mass from $m_0$ to $m$. As we discussed above, this simple model captures the physics of a number of physical systems, ranging from ideal harmonic chains to the low energy properties of interacting fermions and bosons in 1$d$\cite{Cazalilla2011}. 

Before describing qualitatively the main features of the statistics of the work done by quenching the mass in the quadratic field theory of Eqs.~\reff{hamG} and \reff{ham1}, we briefly recall some basic facts about the statistics of the work\cite{Campisi2011,Silva2008,Talkner2007,GS12}. In the case of a quench occurring from an initial state at zero temperature, the probability distribution $P(W)$ of the work $W$  done on the system by the quench is given by 
\eq[-1]{P(W) = \sum_{n\ge 0} \delta(W-E(n)+E_0(0)) |\la n|\Psi_0\ra|^2} 
where $|n\ra$ are the eigenstates of the post-quench Hamiltonian $H$ with energies $E(n)$ while $E_0(0)$ is the energy of the ground state $|\Psi_0\ra \equiv |0\ra_0$ of the pre-quench Hamiltonian $H_0$ with eigenstates $|n\ra_0$ and eigenvalues $E_0(n)$. 
In the specific case of the quench we are interested in $H_0 \equiv H(m_0)$ while $H \equiv H(m)$ with $H$ given in Eq.~\reff{ham1}.
Introducing the generating (or characteristic) function\cite{Talkner2007} 
\eq{G(t) \equiv \langle \rme^{-i Wt} \rangle = \int_{-\infty}^{+\infty} {\rm d}W \; {\rm e}^{-iWt} P(W),}
of the probability distribution $P(W)$ one finds
\eq[0]{G(t)=\la\Psi_0| {\rm e}^{iH_0 t} {\rm e}^{-iHt}|\Psi_0\ra = {\rm e}^{iE_0 t} \la\Psi_0|{\rm e}^{-iHt}|\Psi_0\ra .}
Performing a Wick rotation to imaginary ``time" $t \mapsto -iR$, the amplitude $\la\Psi_0| {\rm e}^{-HR}|\Psi_0\ra$ can be interpreted as
the partition function of the 
classical system corresponding to $H$, in a $(d+1)$-dimensional film of thickness $R$ with boundary states described by $|\Psi_0\ra$ on both boundaries of the film\cite{Gambassi2011}. Partition functions in such geometries are well-known in the study of the so-called \emph{critical Casimir effect}\cite{krech94,*krech99,*Gambassi2009,Gambassi09} and finite-size scaling\cite{fss1,fss2,fss3}. In particular, the corresponding free energy can be decomposed into three contributions according to their scaling upon increasing $R$:
\eq[fe]{\ln{G(R)} = -L^d [f_b R+2f_s+f_C(R)],}
where $L\to\infty$ is the linear size of the system in all $d$ dimensions parallel to the film. Here $f_b \equiv \lim_{L\to\infty} \Delta E_{gs}/L^d$ is the bulk free energy density, where $\Delta E_{gs} \equiv E(0)-E_0(0)$; $f_s$ is the surface free energy associated with each of the two (identical) boundaries of the film and $f_C(R)$ is the remaining finite-size contribution which represents an effective interaction between the two boundaries and decays to zero for large $R$. Upon approaching a critical point of the classical (and therefore quantum) system, this finite-size contribution acquires a universal character encoded in a scaling function of the ratio $R/\xi$ where $\xi$ is the bulk correlation length related to the inverse gap of the quantum system. This universal contribution is known as the critical Casimir effect,
which turned out to influence significantly the behaviour of, inter alia,
soft matter systems and colloids at the micrometer and sub-micrometer scale\cite{krech94,*krech99,*Gambassi2009,Hertlein08,Gambassi09}. 
In principle, $G(t)$ in Eq.~\reff{0} can be calculated by analytic continuation of Eq.~\reff{fe} to real ``times" $R=it$,
\eq[fe1]{\ln{G(t)} = -L^d[i f_b t +2f_s+f_C(it)],}
though particular care might be required 
in some cases for quenches across a quantum phase transition\cite{Heyl2012b}. 
Accordingly, when the post-quench Hamiltonian is close to a critical point, $G(t\to +\infty)$ --- and therefore $P(W)$ --- is expected to acquire universal features which can be inferred from the critical Casimir effect. The work probability distribution is then obtained via the inverse Fourier transform
\eq[0a]{P(W) =  \int_{-\infty}^{+\infty} \!\!\frac{{\rm d}t}{2\pi} \; {\rm e}^{+iWt} G(t) }
of $G(t)$.

In the case of a quantum quench originating from a thermal initial state at a temperature $\beta^{-1}$, $P(W)$ is given by a generalization of Eq.~(\ref{-1}):
\eq{P(W) = \sum_{n,n_0} \delta(W-E(n)+E_0(n_0)) \; p_{n_0} \; |\la n|n_0\ra_0|^2 ,}
where the occupation probabilities $p_{n}$ of the pre-quench energy eigenstates are $p_{n} = {\rm e}^{-\beta E_0(n)}/\sum_{n} {\rm e}^{-\beta E_0(n)}$ and $|n\ra_0$ are the eigenstates of $H_0$ with energies $E_0(n)$. The corresponding generating function is therefore given by
\eq[1a]{ G(t) = \frac{\text{Tr}\{ {\rm e}^{-\beta H_0} { {\rm e}^{iH_0 t}} {\rm e}^{-iHt} \}}{\text{Tr}\{ {\rm e}^{-\beta H_0} \}} .}
We refer to this quench originating from a thermal initial state as \emph{thermal} quantum quench, in contrast to the case with zero initial temperature, which we refer to as \emph{ordinary} quantum quench.

\begin{figure}[h!]
\centering
\includegraphics[width=.95\columnwidth]{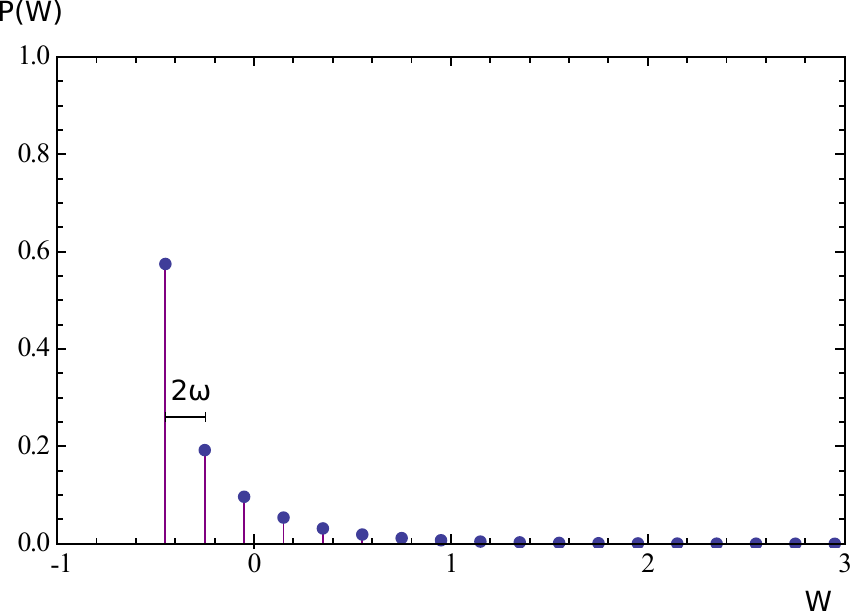}
\caption{\label{sho} \small Probability distribution $P(W)$ of the work $W$ done
on a single harmonic oscillator by a quench of its characteristic frequency from $\o_0=1$ to $\o=0.1$.
Vertical lines indicate $\delta$-functions, the effective strength of which is represented in the vertical axis. }
\end{figure}
\begin{figure*}[h!bt] 
\centering
\includegraphics[width=.8\textwidth]{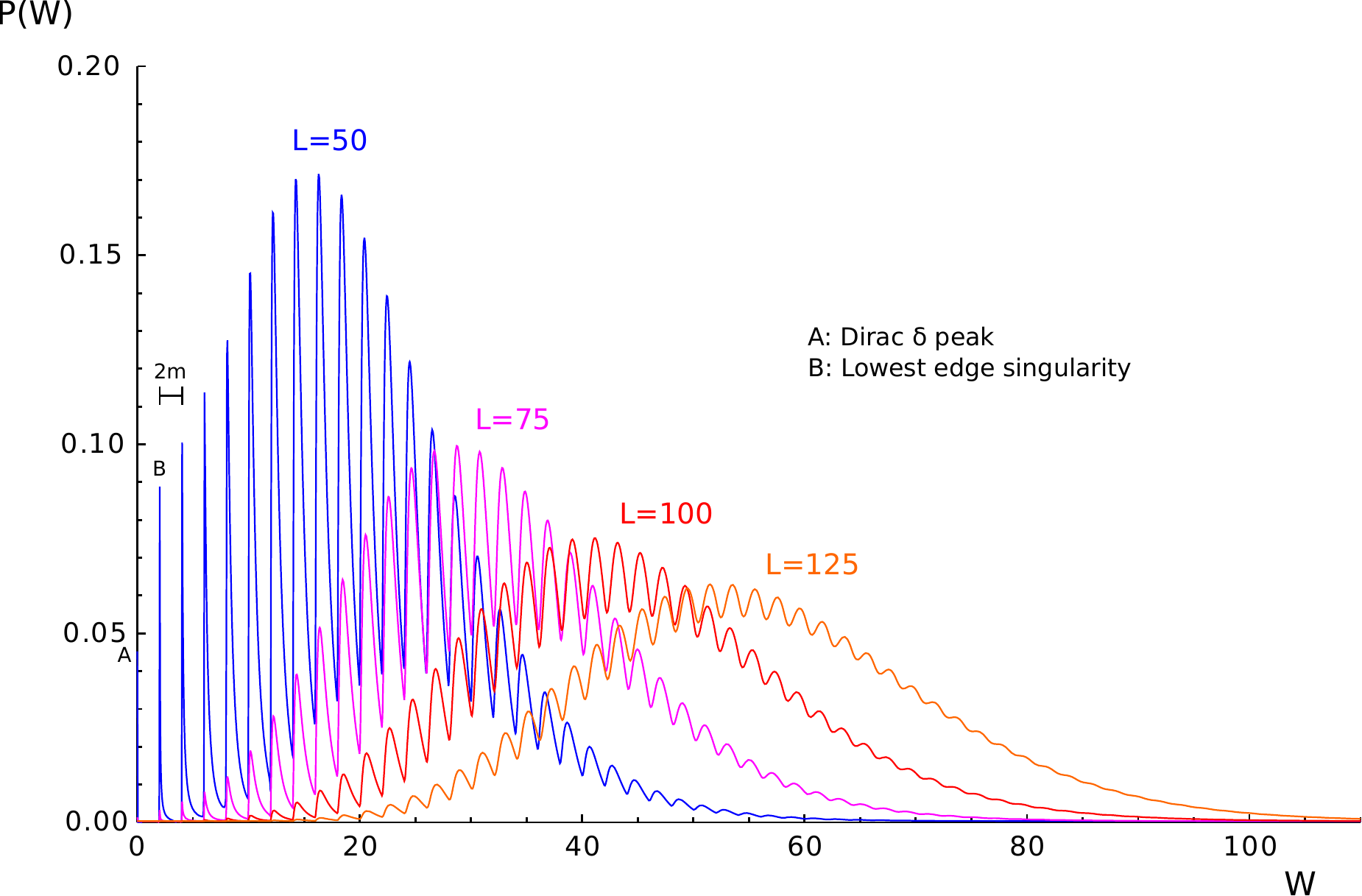} 
\caption{\label{fig1} \small 
Probability density $P(W)$ of the extensive work $W$ done by an ordinary quantum quench in a system of $1d$ non-interacting bosons with $m=1$,  
$m_0=0.1$ and increasing system size $L=50$, 75, 100, and 125. Sharp edge-singularity peaks appear at even integer multiples of $m$, superimposed on a wider bell-shaped distribution the mean and width of which increase upon increasing $L$. In particular, A indicates a contribution $\propto \delta(W)$ associated with the transitions between the pre- and post-quench ground states, whereas B indicates the lowest edge singularity, the shape of which is investigated here.}
\end{figure*}
The general features of the probability distribution $P(W)$ associated with an ordinary quantum quench can be deduced already from Eqs.~\reff{-1} and \reff{fe1}. It is instructive to consider first the case of a single quantum harmonic oscillator after a quench of its frequency from $\o_0$ to $\o$. The associated $P(W)$ consists of a sequence of equidistant Dirac $\delta$-peaks corresponding to all the allowed transitions from the initial ground state to the eigenstates of the post-quench Hamiltonian, as shown in Fig.~\ref{sho} for a specific choice of the values of $\omega_0$ and $\omega$. According to Eq.~\reff{-1}, the effective strength of each peak is set by the transition probability between the states which are involved in the transitions. Because of the behaviour of the wavefunctions of the harmonic oscillator under spatial inversion, only transitions to even excited states are allowed from the ground state (which is invariant under inversion) and consequently the spacing between consecutive peaks is given by $2\omega$.
The $\delta$-peak at the lowest frequency corresponds to the transition to the new ground state, since this requires the smallest possible amount of work.

For the system of non-interacting (massive) bosons in Eq.~\reff{ham1} the transitions to excited states lead to the creation of pairs of particles which, by conservation of momentum, must have opposite momenta $k$ and $-k$ with energies $\omega_{-k}(m)=\omega_k(m)\ge m$. Therefore each of the previous $\delta$-peaks (except for the lowest one that corresponds to the transition from the pre- to the post-quench ground state) is now replaced by a continuous distribution with a lower threshold corresponding to the minimum energy cost for the creation of $n$ pairs of particles with mass $m$, i.e., $2nm$. This is clearly visible in Fig.~\ref{fig1}, where we report $P(W)$ for a given quench of the mass $m$ of the model~\reff{ham1} in one spatial dimension, for various value of the size $L$ of the chain. Given that $P(W)$ vanishes identically below the extensive threshold $\Delta E_{gs} \simeq f_b L^d$, in Fig.~\ref{fig1} and in what follows we refer $W$ to this value, so that the transition between the ground states of the pre- and post-quench Hamiltonian give rise to a $\delta$-peak located at $W=0$.
In passing we observe that $\Delta E_{gs}$ is nothing but the work $W_{\rm rev}$ that one would do at zero temperature $\beta^{-1}=0$ on the system during a transformation which changes the parameter $m$ from its pre- to its post-quench value and which is reversible in the sense of thermodynamics. In fact, this reversible work   $W_{\rm rev}$ is equal to the difference $\Delta F = F_\beta(m) - F_\beta(m_0)$, where  $F_\beta(m)$ is the free energy of the system in contact with a thermal bath at temperature $\beta^{-1}$ and Hamiltonian $H(m)$, which renders the ground-state energy for $\beta\to\infty$. Accordingly, the mean of the quantity $W - \Delta E_{gs} = W - W_{\rm rev}$ studied below is nothing but the so-called irreversible work  
$W_{\rm irr}$\cite{A} (see, e.g., Ref.~\onlinecite{Dorner2012} where $W_{\rm irr}$ and the associated irreversible entropy production $\Delta S_{\rm irr}$ are studied for a thermal quench of the Ising chain).

\red{In an interacting bosonic system, the pattern described above would no longer consist of equidistant peaks and it may involve additional excitations that are not pairs of quasi-particles with opposite momenta (as these properties are specific to non-interacting systems) but also bound states or multi-particle excitations. Bound states result in $\delta$-peaks in $P(W)$ located at the bound state energy, while multi-particle excitations result in peaks with a lower threshold equal to the sum of the masses of the quasi-particles created. %
For example, in the sine-Gordon model the excitations consist of solitons, antisolitons, and their bound states, referred to as breathers, the number of which increases as $\beta$ decreases \cite{ZamosG}. For $\beta\to 0$ the breathers have approximately equidistant masses, corresponding to the energy levels of the Gaussian approximation discussed here, and their number increases without bounds. 
If the initial state $|\Psi_0\rangle$ of the dynamics corresponds to a so-called boundary integrable state \cite{Ghoshal_94} for the 2$d$ film (strip) obtained in imaginary time, the leading effect of a non-vanishing interaction is the emergence of $\delta$-peaks in the plot of $P(W)$ as a function of $W$, corresponding to the creation of breathers, which is accompanied by a progressive leftward shift of the positions of the pre-existing peaks due to the non-parabolic shape of the potential \cite{Gritsev2007}. 
Interactions which break integrability but do not result into a significant production of multi-particle excitations are expected (on the basis of the form-factor perturbation
theory \cite{DMS}) to affect solely the positions of the
peaks. For stronger perturbations, instead, multi-particle excitations emerge in the initial state which cause additional peaks with lower thresholds to appear not only at even but also at odd multiples of the quasi-particle masses. The same occurs if the post-quench Hamiltonian
is integrable but the initial state does not correspond to a boundary integrable state.}

\begin{figure}[h!]
\centering
\includegraphics[width=.97\columnwidth]{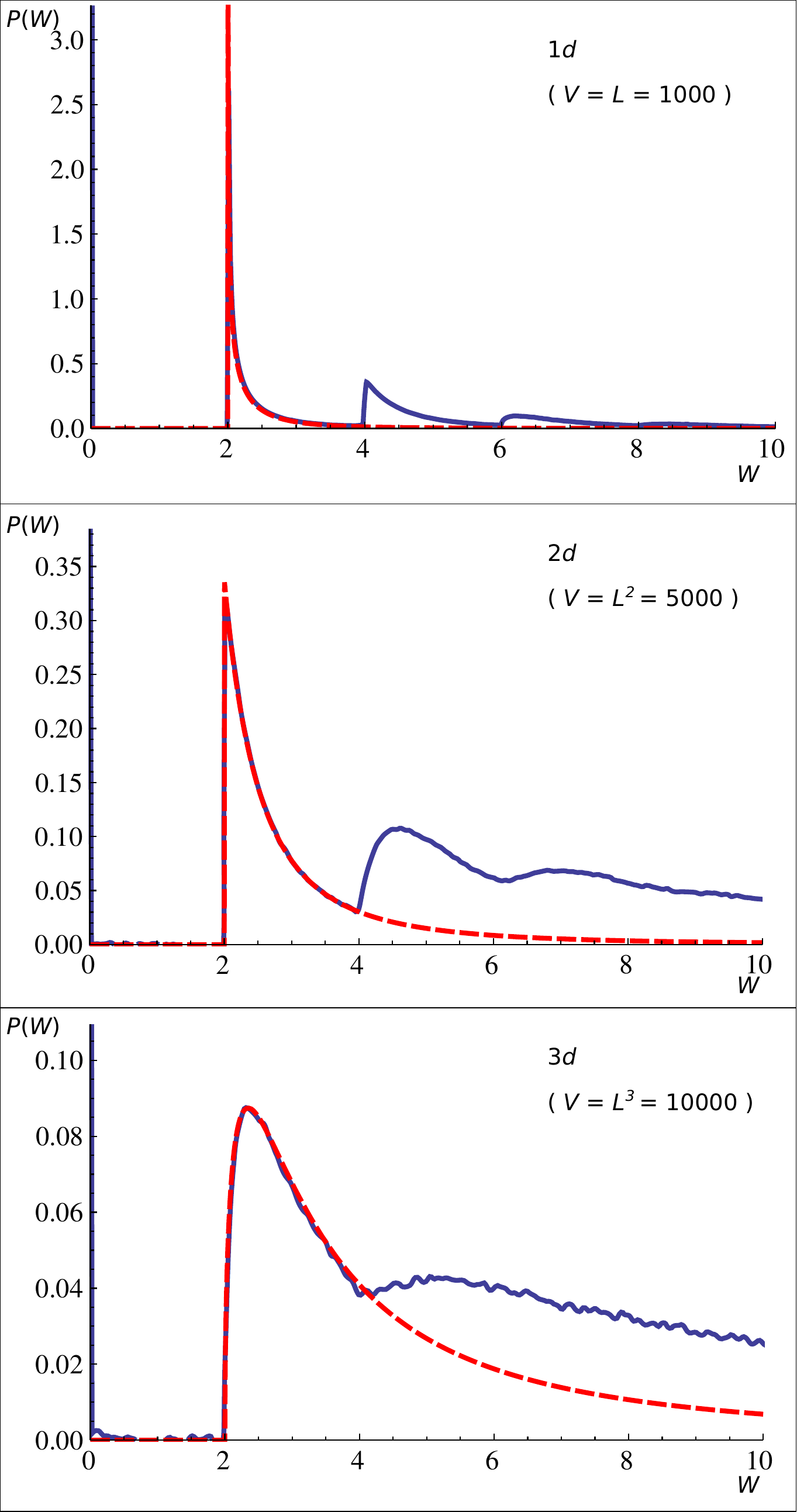} 
\caption{\label{fig1b} \small Lowest edge singularity of $P(W)$ for a quench in the non-interacting bosonic system in Eq.~\reff{ham1} with  $m=1$ and $m_0=0.8$ and, from top to bottom, spatial dimensionality $d=1$, 2 and 3. The solid (blue) lines are the numerical evaluation of Eq.~(\ref{0a}) with Eq.~(\ref{qqpf}), while the dashed (red) lines correspond to the analytic expressions in, c.f., Eqs.~(\ref{esth}) and (\ref{chi}). Here $W$ is measured in units of $m$ and, for convenience, different values of $L$ have been used in each plot. Note that, depending on the dimensionality $d$, $P(W)$ diverges ($d<2$), tends to a finite value ($d=2$) or vanishes ($d>2$) upon approaching the threshold $W=2m$. 
The agreement between the numerical and the analytical curves is excellent for $W<4m$, where the analytic expression is indeed expected to reproduce the numerical data, as discussed further below in section~\ref{sec:3}.
The noise in the blue curves, which becomes increasingly relevant 
as the dimensionality increases, is an artifact of the numerical integration.}
\end{figure}

\red{However, some of the features of $P(W)$ mentioned above for non-interacting systems are robust and unaffected by the inclusion of interactions.} In particular, the low-energy part of $P(W)$ always consists of a $\delta$-peak at $W=0$, with an amplitude given by the square ${\mathcal F}$ of the so-called fidelity $|\la 0|\Psi_0\ra|$ between the pre- and post-quench ground state which, from Eq.~\reff{fe1}, is 
\eq[eq:fid]{\mathcal{F} \equiv |\la 0|\Psi_0\ra|^2= {\rm e}^{-2L^d f_s}.} 
In addition, provided the system has a gap $m\neq 0$, a lower threshold emerges at $W= 2m$ which reflects the fact that the lowest allowed excitations consist of pairs of quasiparticles with opposite momenta, each of which has energy at least equal to the energy gap $m$. 
Just above this threshold, $P(W)$ turns out to exhibit an \emph{edge singularity} whose profile is determined by the way $f_C$ in Eq.~(\ref{fe}) decays for 
$R\to +\infty$.
Depending on the space dimensionality $d$ of the system, this edge singularity (already present in the non-interacting case) displays different qualitative features, as shown in Fig.~\ref{fig1b} for an instance of quench of the model in Eq.~\reff{ham1} and $d=1$, 2, and $3$ (from top to bottom). Further below we rationalize these features in full generality, by exploiting the connection with the critical Casimir effect.
%
%
%%%%%%%%%%%%%%%%%%%%%%%%%%%%%%%%%%%%%%%%%%%%%%%%%%%%%%%%%%%%%%%%%
%%%%%%%%%%%%%%%%%%%%%%%%%%%%%%%%%%%%%%%%%%%%%%%%%%%%%%%%%%%%%%%%%
%
%

\begin{figure}[h!b] 
\centering
\includegraphics[width=.95\columnwidth]{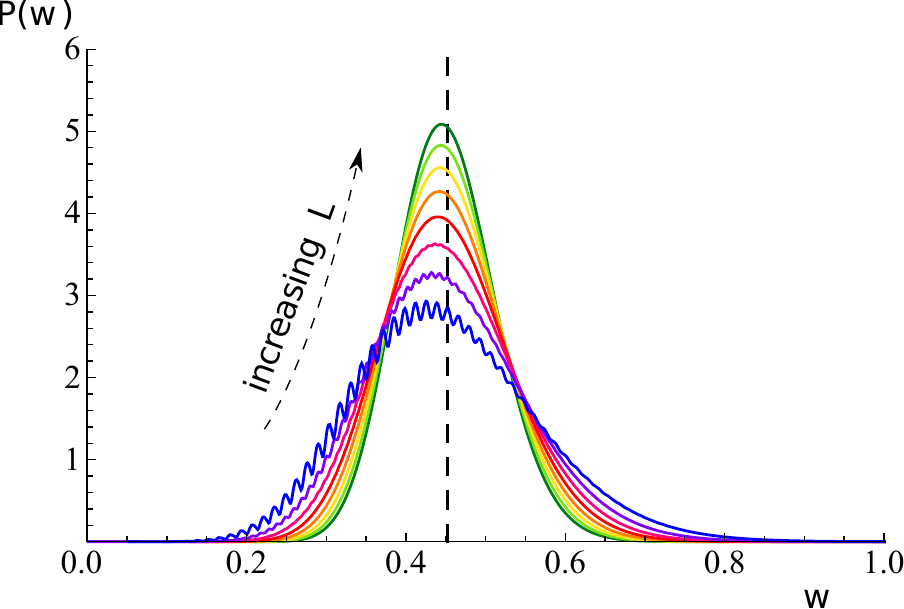}
\caption{\label{fig1c} \small Probability density $P(w)$ of the intensive work $w=W/L^d$ done on the one-dimensional ($d=1$) bosonic model in Eq.~\reff{ham1} with $m=1$, $m_0=0.1$, and of size $L$ increasing from 150 to 500 in steps of 50. For large $L$ the prominent feature is a bell-shaped distribution which is completely described by its mean value, standard deviation and skewness (in order of decreasing importance) and which approaches a  as $L$ increases. The dashed vertical line indicates the asymptotic value of the mean for $L\to \infty$.
}
\end{figure}

The fine structure of overlapping peaks (described above for the non-interacting case) fades out, however, upon increasing the system size $L$ and $P(W)$ gradually develops a highly peaked Gaussian distribution with an extensive mean value $\propto L^d$ and a relative width $\propto L^{-d/2}$ which vanishes as $L\to +\infty$. As we show rigorously further below, this is a direct consequence of the fact that $\ln{G(t)}$ is extensive, i.e., proportional to $L^d$, as can be seen from Eq.~(\ref{fe1}). 
The emergence of a Gaussian distribution as $L$ increases is shown in Fig.~\ref{fig1c} where, for the same system and quench as in Fig.~\ref{fig1}, we report the distribution $P(w)$ of the intensive work $w\equiv W/L^d$ for increasing values of $L$. While typical fluctuations around the mean 
(and most probable) value of $w$ are indeed characterized by a Gaussian distribution as $L\to\infty$, large fluctuations display non-trivial and universal features\cite{GS12}.
%

%
%
%%%%%%%%%%%%%%%%%%%%%%%%%%%%%%%%%%%%%%%%%%
%%%%%%%%%%%%%%%%%%%%%%%%%%%%%%%%%%%%%%%%%%
%
%

For a thermal quantum quench, thermal excitations in all energy levels pre-exist in the initial state and they do not form only pairs of opposite momenta as in the case of the ordinary quench discussed above.
Thus additional peaks appear in $P(W)$ corresponding to transitions from all the energy levels of the pre-quench Hamiltonian to all those of the post-quench one, with an amplitude which decays exponentially upon increasing the corresponding energy difference. These peaks of thermal origin are present also below the lower threshold  
--- which is a characteristic feature of the quench originating from a zero-temperature initial state (i.e., the ground state, or more generally, a generic eigenstate of the pre-quench Hamiltonian) --- and they obscure the zero-temperature structure of $P(W)$. In addition, for a thermal quench, the relation of $P(W)$ with the statistical mechanics of a classical system in a film encoded in Eq.~\reff{fe1} is no longer valid.

\section{Ordinary quantum quench}\label{sec:2}

In this section we show explicitly how the qualitative features mentioned above emerge in the case of the non-interacting bosonic model of Eq.~(\ref{ham1}) and we discuss in detail the scaling properties of $P(W)$ as a function of the system size $L$. The calculation of $G(t)$ for quenches of the mass in Eq.~(\ref{ham1}) is clearly simplified by the non-interacting nature of the problem. The initial state $|\Psi_0\ra$ can be decomposed on the basis of eigenstates of the post-quench Hamiltonian $H$ (generated by the action of the post-quench creation operators $a_k^\dg$ on the vacuum $|0\ra$ of $H$) as a set of ``Cooper pair'' excitations; more specifically (see appendix~\ref{app}):
\eq[inst]{|\Psi_0\ra = \prod_k(1-\lambda_k^2)^{1/4} \; \exp\(-\int_k \frac{\lambda_k}{2} a^\dg_{-k} a^\dg_k\)|0\ra ,}
where
\eq[eq:lambdak]{\lambda_k = \frac{\o_{0k}-\o_k}{\o_{0k}+\o_k}.}
Here and in what follows,
$\o_{0k}\equiv \o_k(m_0)$ and $\o_k\equiv \o_k(m)$ indicate the dispersion relations of the pre- and post-quench Hamiltonian, respectively. 
This very special type of state is a so-called \emph{squeezed vacuum} state and consists solely of pairs of particles of opposite momenta. As discussed in appendix~\ref{app}, it emerges because the pre- and post-quench creation and annihilation operators are related by a linear Bogoliubov transformation. These transformations and the resulting squeezed states characterize a large number of quantum quenches in non-interacting models, both bosonic and fermionic, while in interacting systems (that cannot be equivalently described by non-interacting ones) they are not of such generic utility but they emerge, instead, only in special cases of quenches\cite{FM10,SFM12}.

Due to the non-interacting nature of the model, the generating function $G(t)$ of the statistics of the work is the product of those $\{G_k(t)\}_k$ corresponding to each mode with momentum $k$, i.e., $G(t)=\prod_k G_k(t)$. The function $G_k(t)$ for a single harmonic mode is calculated in Eq.~(\ref{qqpf0}) of appendix~\ref{app} and is given by
\eq[qqptext]{G_k(t) = {\rm e}^{i(\o_{0 k}-\o_k)t/2}\;\sqrt{ \frac{1-\lambda_k^2}{1-\lambda_k^2 {\rm e}^{-2i\o_k t}}} .}
Accordingly, for $L\to+\infty$,
\eqa{  \frac{\ln G(t)}{L^d} &= -i f_b
\; t + \frac{1}{2}\int_k \ln\(1-\lambda_k^2\) \nonumber \\
 &- \frac{1}{2} \int_k \ln\(1-\lambda_k^2 {\rm e}^{-2i\o_k t}\) \label{qqpf},}
where $f_b = \lim_{L\to\infty} 
\Delta E_{gs}/L^d = \int_k (\o_k-\o_{0k})/2$. 
The first term on the r.h.s. can be omitted if, consistently with the conventions introduced above, we rename $W$ as $W+\Delta E_{gs}$ so that 
we measure $W$ starting from $\Delta E_{gs}$. 
Note that the integrals involved in the second and third term on the r.h.s.~of Eq.~(\ref{qqpf}), with $\o_k=\sqrt{k^2+m^2}$, are separately convergent in the ultraviolet (UV) $k\to\infty$ provided that $d<4$ because $\lambda_k \propto k^{-2}$ in the same limit. 
By comparing Eq.~\reff{qqpf}  with the decomposition in Eq.~\reff{fe1}, one identifies the second term on the r.h.s. with $-2 f_s$ while the third one turns out to be the analytic continuation of the critical Casimir contribution $-f_C(R)$ in Eq.~\reff{fe} to imaginary film thickness $R \to it$. 
We consider this issue in more detail in section.~\ref{sec:3}

The cumulants of $P(W)$, their scaling with the system size $L$, the fidelity ${\cal F}^{1/2}$, and the profile of the lowest edge singularity can be derived from Eq.~\reff{qqpf}. 
In particular, the cumulants ${\kappa_n}_W$ can be calculated from $\ln G(t)$ using the relations
\eq{{\kappa_n}_W = i^n \left.\frac{{\rm d}^n}{{\rm d} t^n} \ln G(t) \right|_{t=0}.}
The resulting expressions for ${\kappa_n}_W$ involve an integration over the momentum $k$, which we have evaluated in the continuum limit, i.e.,  $\int_k \equiv \Omega_d/(2\pi)^d \int_0^\infty dk \; k^{d-1}$ whenever the corresponding integral with a large-momentum cutoff $\Lambda$ (related to the inverse lattice spacing on a lattice) converges for $\Lambda \to \infty$. Depending on the space dimensionality one finds for the mean work 
(i.e., the so-called irreversible work\cite{A}):
\eqa{ & \mu_W \equiv {\kappa_1}_W = L^d   \int_k \frac{(m_0^2 - m^2)^2}{4 \o_{0k} (\o_k+\o_{0k})^2} = \frac{L^d}{(2\pi)^d} \times \nn \\
& 
\begin{cases} 
[m_0^2-m^2+2m^2 \ln(m/m_0)]/4 			&\text{ in } 1d, \\
(m_0-m)^2(m_0+2m)\pi/6 				&\text{ in } 2d, \\
[(m^2-3m_0^2)(m_0^2-m^2)/4 - m^4 \ln(m/m_0)	& \\
\quad + (m_0^2-m^2)^2 \ln(2\Lambda/m_0)]\pi/4 	&\text{ in } 3d. \\
\end{cases} \label{k1}}
Notice that $\mu_W$ is UV-convergent in $d=1$  and 2 
and it is generically finite upon quenching to the critical point $m=0$. On the other hand, it is finite for $d=2$ and 3, when the initial state approaches a critical one $m_0=0$, while it diverges in the same limit for $d=1$.
Note that the mean work $\la W \ra$ can be expressed as $\la \Psi_0 | H(m)-H(m_0)|\Psi_0\ra $, i.e., from Eq.~\reff{ham1}, $\la W \ra = (m^2-m_0^2) \la \Psi_0 |\int_k \phi_k\phi_{-k}/2|\Psi_0\ra$, where this latter expectation value depends only on $m_0$ and therefore $\la W \ra$ depends linearly on $m^2$: the additional dependence of $\mu_W$ on $m$ displayed in Eq.~\reff{k1} comes about because $W$ is measured here with respect to $\Delta E_{gs} = E(0)- E_0(0)$ (i.e., $\mu_W$ is actually the mean irreversible work) and $E(0)$ depends on $m^2$ in a non-linear fashion. Further below we will extend this argument to a thermal quantum quench in order to argue that $\la W \ra$ depends linearly on $m^2$ also in that case.

The variance is
\eqa{ \sigma_W^2 & \equiv {\kappa_2}_W = L^d   \int_k \frac{(m_0^2 - m^2)^2}{8 \o^2_{0k}}  \nn \\
&= \frac{L^d}{(2\pi)^d} (m_0^2-m^2)^2 \times
\begin{cases} 
\pi/(8m_0) 					&\text{ in } 1d, \\
-\ln(m_0/\Lambda)\pi/4 			&\text{ in } 2d, \\
-m_0\pi^2/4 + \pi \Lambda/2 			&\text{ in } 3d. \\
\end{cases} \label{k2}}
$\sigma_W^2$ is UV-convergent only in $d=1$. Note that the relative standard deviation is
\eq{ \frac{\sigma_W}{\mu_W} \propto L^{-d/2} }
i.e., the relative fluctuations of $W$ vanish for $L\to+\infty$. 
Analogously, from Eq.~(\ref{qqpf}) one finds
\eqa{ {\kappa_3}_W & = L^d (m_0^2 - m^2)^2  \int_k \frac{\o_k^2+\o_{0k}^2}{8 \o^3_{0k}} = \frac{L^d}{(2\pi)^d} (m_0^2 - m^2)^2 \times \nn \\ 
& 
\begin{cases} 
[(m/m_0)^2 -1 + 2\ln(2\Lambda/m_0)]/4 					&\text{ in } 1d, \\
(m^2-3m_0^2)\pi/(4m_0) + \pi \Lambda/2					&\text{ in } 2d, \\
(3m_0^2-2m^2)\pi/4 + & \\
\quad (m^2-2m_0^2)(\pi/2)\ln(2\Lambda/m_0) + \pi\Lambda^2/2 		&\text{ in } 3d. \\
\end{cases} \label{k3}}
The skewness  of $P(W)$ is defined as
\eq{ {\gamma_1}_W \equiv \frac{{\kappa_3}_W}{\sigma_W^3} \propto L^{-d/2} ,}
i.e., it vanishes for $L\to+\infty$.
As heuristically expected on the basis of the central limit theorem, the expressions above show that the mean work scales proportionally to the system volume $L^d$, while the relative variance and skewness are proportional to the inverse square root of the volume, indicating that the probability distribution $P(W)$ of the intensive work $w$ (or, alternatively, of the normalized work $W/\langle W\rangle$) becomes asymptotically sharply peaked as $L$ increases. In fact, the scaling of the distribution $P(W)$ for large $L$ can be easily derived: according to Eq.~(\ref{qqpf}), $\ln G(t)$ is proportional to $L^d$, i.e., it is extensive and therefore each of the cumulants ${\kappa_n}_W$ is extensive too. It is more interesting to study the probability distribution of the work per unit volume (intensive work) $w = W/L^d$ whose characteristic function is $\tilde{G}(t) = G(t/L^d)$. As long as one is interested in small relative fluctuations of the intensive work around its mean, i.e., in the behaviour of $P(W)$ close the central peak (see Fig.~\ref{fig1c}), it is possible to expand the function $\ln G(t)/L^d$ around $t=0$:
\eq{\ln G(t)/L^d = 1 - i a t - \tfrac{1}{2} b t^2 + \tfrac{1}{3!} i c t^3 + \mathcal{O}(t^4)}
where the constants $a,b,c$ become independent of $L$ for $L$ large and are given by $a = \mu_W/L^d , \, b = \sigma^2_W/L^d $ and $c = {\kappa_3}_W/L^d $. 
Accordingly, the asymptotic form of $\tilde{G}(t)$ is
\eqa{ \tilde{G}(t) = & \exp\[-iat - \frac{1}{2 L^d} (b-a^2) t^2 + \right. \nn \\
& \left. + i \frac{1}{3! L^{2d}} (c+2a^3-3ab) t^3 + \mathcal{O}\(\frac{t^4}{L^{4d}}\)\],}
which is governed by the lowest-order cumulants. More specifically, the probability distribution of $w$ for large $L$ is  with mean value $a$ and variance $(b-a^2)/L^d$ which vanishes as $L\to+\infty$. Figure~\ref{fig1} shows typical distributions of the work $P(W)$ for various values of $L$ while Fig.~\ref{fig1c} displays the asymptotic shape of $P(w)$ as $L$ grows.

\begin{figure}[htbp]
\centering
\includegraphics[width=\columnwidth]{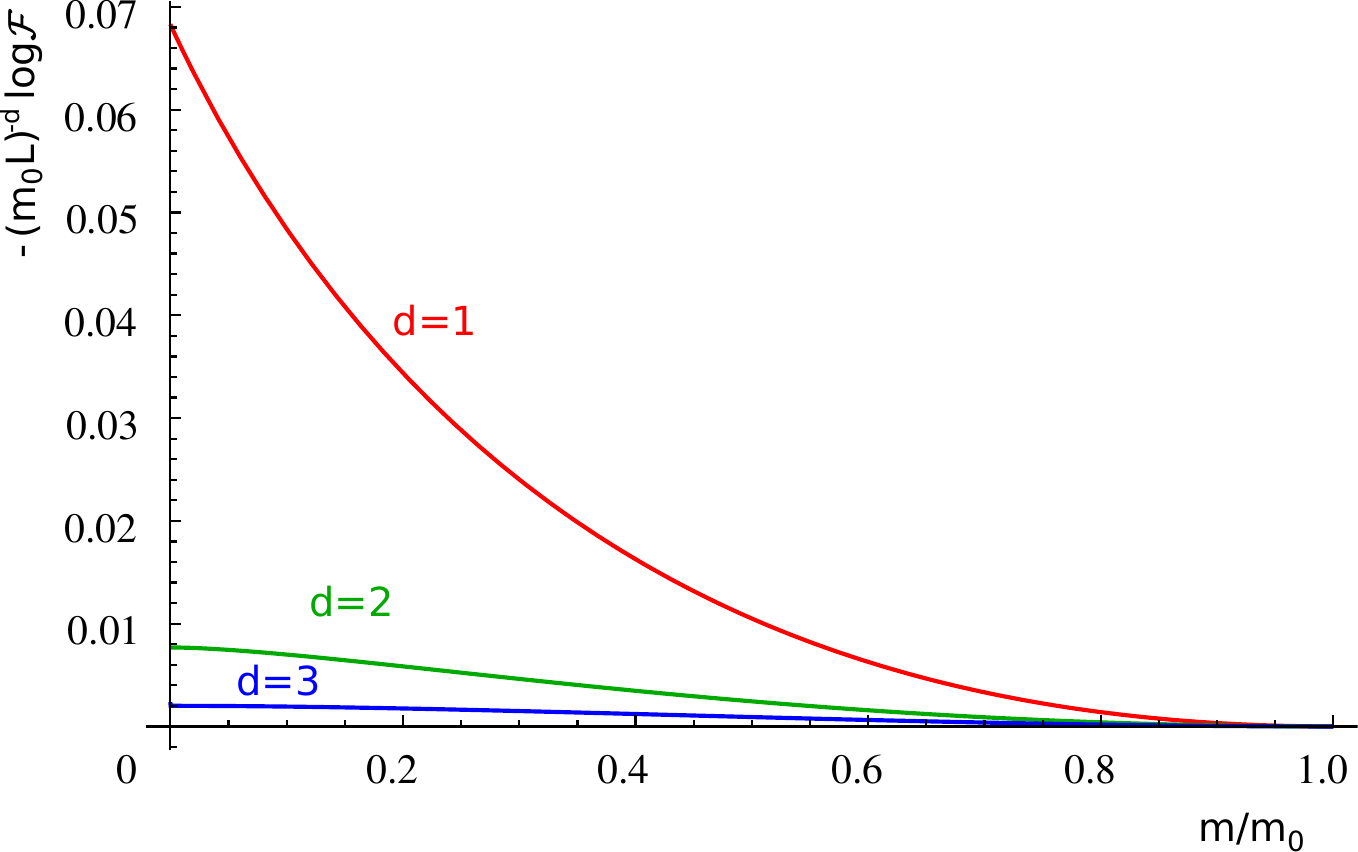}
\caption{\label{fidel} \small Scaling plot of the square fidelity $\mathcal F$ (see Eqs.~\reff{eq:fid} and \reff{fid}). The dimensionless quantity $ -(m_0 L)^{-d} \ln\mathcal{F} $ is plotted as a function of $m/m_0<1$ in $d=1$ (red), 2 (green) and 3 (blue curve). Due to the symmetry of $\mathcal{F}$ under the exchange $m\leftrightarrow m_0$, the behaviour for $m/m_0>1$ can be readily inferred from the present plot.}
\end{figure}

In order to characterize the behaviour of the probability $P(W)$ it is useful to calculate the square fidelity $\mathcal{F}$ between the pre- and post-quench ground state. Recalling its definition in Eq.~\reff{eq:fid}, ${\cal F}$ is given by the exponential of the second contribution in Eq.~(\ref{qqpf}), i.e.,
\eq[fid]{ \mathcal{F}(m_0,m) = \exp\[- L^d \int_k \ln \(\frac{\o_k+\o_{0k}}{2 \sqrt{\o_{0k} \o_k}}\)\].}
Note that $\mathcal{F}(m_0,m) = \mathcal{F}(m,m_0)$.
The integral is UV-convergent for $d<4$, in which case it can be expressed in terms of elliptic integrals and $-\ln\mathcal F$ takes a scaling form of $m$ and $m_0$. In particular, $-(m_0 L)^{-d} \ln\mathcal F$ is the function of $m/m_0$ plotted in Fig.~\ref{fidel} for various values of $d$ and which, for $m=0$ (or, equivalently, for $m_0\to\infty$) takes the values
\eq[F:crit]{ \ln\mathcal{F}(m_0,0) = -\(\frac{m_0L}{2\pi}\)^d\! \times \!\begin{cases}
2(1-\pi/4) 					&\text{ in } 1d, \\
\pi(2\ln2-1)/4 				&\text{ in } 2d, \\
\pi(3 \pi-8)/9 			&\text{ in } 3d. \\
\end{cases}}
For later convenience, we report here explicitly the behaviour of ${\cal F}(m_0,m)$ in the two limiting cases for the value of the masses $m$ and $m_0$:
\eq[F:scal:asy]{{\cal F}(m_0,m) \simeq \begin{cases}
\rme^{- (m_0L)^d C_d} & \mbox{for}\quad m_0 \gg m,\\
\rme^{-(mL)^d C_d} & \mbox{for}\quad m_0 \ll m,
\end{cases}} 
where $C_d$ is the value for $m/m_0=0$ of the function plotted in Fig.~\ref{fidel}, which can also be inferred from Eq.~\reff{F:crit}.
%
%
%
%%%%%%%%%%%%%%%%%%%%%%%%%%%%%%%%%%%%
%%%%%%%%%%%%%%%%%%%%%%%%%%%%%%%%%%%%

As we have anticipated, some of the expressions in Eqs.~(\ref{k1}), (\ref{k2}), (\ref{k3}), and (\ref{fid}) for the first three cumulants and the fidelity, respectively, or their generalized susceptibilities (i.e., their derivatives with respect to $m$ or $m_0$) diverge for $m\to 0$ and/or $m_0\to 0$, especially for $d=1$. 
These infrared (IR) divergences indicate the emergence of a universal scaling behaviour of the corresponding quantities in the neighbourhood of critical points. In particular for $m\to 0$, the mean work susceptibility $\partial^2 \mu_W/\partial m^2\propto \ln(m/m_0) $ exhibits a logarithmic IR-divergence in $1d$ for $m$ or $m_0\to 0$ (see Eq.~\reff{k1}).
Analogously, the fidelity ${\mathcal F}^{1/2}$ develops a non-analytic behaviour upon approaching the critical point $m=0$. Indicating by $g$ the linear coupling of the most relevant operator which drives the system away from the critical point ($\phi^2$ in the present case --- see Eq.~\reff{hamG} --- and therefore $g=m^2$ and $g_0=m_0^2$), one defines generalized fidelity susceptibilities as $\chi_n(g_0,g) \equiv - L^{-d} \partial_g^n \ln {\mathcal F}^{1/2}(g_0,g)$~\cite{CamposVenuti2007,DeGrandi2010}.
In particular, the fidelity susceptibility $\chi_2(g,g)$ is expected to scale as $\chi_2(g,g) \simeq |g-g_c|^{\nu d-2}$ for $g\to g_c$ (see, e.g., Ref.~\onlinecite{CamposVenuti2007,DeGrandi2010,GS12}), where $g_c$ is the critical value of the coupling $g$ and $\nu$ is the bulk critical exponent which controls the divergence of the correlation length $\xi(g\to g_c) \simeq |g-g_c|^{-\nu}$ of the system. In the present case $g_c=0$ and, as expected for a  theory, $\nu=1/2$. A direct calculation from Eq.~\reff{fid} shows that $\chi_2(m^2,m^2) \propto (m^2)^{d/2-2}$ for $m\to 0$ and $d<4$, in agreement with the expected scaling.
For $m_0\to 0$, instead, IR-divergences are even more common and appear also at the level of the cumulants themselves, as one can easily see from Eqs.~\reff{k1}, \reff{k2}, and \reff{k3}. 
A physical interpretation for this is that the quadratic potential of the pre-quench Hamiltonian is vanishingly small for the modes with low wavevectors $k$ (as $\omega_{k\to 0}(m_0=0)\simeq k$), the wavefunction of which is therefore extended in space, such that an infinite amount of work is required in order to raise the potential to its post-quench form with $\omega_{k\to 0}(m) = m$. This fact has an important consequence also for the statistics of the large deviations of the work $W$ from its mean (and typical) value, and leads to the occurrence of a condensation transition\cite{GS12}.

\section{The edge singularity and its relation to the critical Casimir effect}\label{sec:3}

The behaviour of $P(W)$ for $W$ close to the threshold is determined by the large-$t$ behaviour of 
$G(t)$, which, in turn, is given by the third contribution in Eq.~(\ref{qqpf}):
\eq[comp1]{ -\frac{1}{2} \int_k \ln(1-\lambda_k^2 {\rm e}^{-2i\o_k t}) \equiv \varphi(t) .}
In order to calculate this asymptotic behaviour, we first expand the logarithm as a Taylor series, recalling that $|\lambda_k| < 1$ for all finite values of $m,m_0$ so that the expansion is always justified. Next we interchange the order of momentum integration and summation over $n$, since, recalling that $\lambda_k\propto k^{-2}$ as $k\to+\infty$, all intermediate expressions are UV-convergent at least as long as $d<4$, and therefore
\eq[series]{ \varphi(t) = -\frac{1}{2} \sum_{n=1}^\infty \frac{1}{n} \int_k \lambda_k^{2n} {\rm e}^{-2 n i \omega_k t} .}
If $m\neq 0$, the stationary phase approximation can be used in order to determine the behaviour for $m t\gg 1$ of each integral in the series 
with $\omega_k \simeq m + k^2/(2m)$, which therefore decays as
\eq[asympt]{\int_k \lambda_k^{2n} {\rm e}^{-2ni\o_k t} \simeq \lambda_0^{2n} \; {\rm e}^{-2nimt} \(\frac{m}{4\pi nit}\)^{d/2},}
with $\lambda_0=(m_0-m)/(m_0+m)$. The series in Eq.~\reff{series} can then be summed up and it gives
\eq[asymptphi]{ \varphi(t) \simeq -\frac12 \(\frac{m}{4\pi it}\)^{d/2} \text{Li}_{1+d/2}(\lambda_0^2 {\rm e}^{-2imt}) ,}
where $\text{Li}_s(x)$ is the polylogarithm function of order $s$ (see, e.g., \S25.12 in Ref.~\onlinecite{HMF}).

Recalling from Eqs.~\reff{0a},  \reff{qqpf}, \reff{eq:fid}, the discussion after Eq.~\reff{qqpf}, and Eq.~\reff{comp1}, that the work distribution $P(W)$ is given by
\eq[PWdisc]{P(W) = \mathcal{F}(m_0,m) \int \frac{{\rm d}t}{2\pi} \, {\rm e}^{iWt - L^d \varphi(t)} ,}
one can see (by formally expanding the exponential in $\varphi$) that each term in the Fourier series of the $(\pi/m)$-periodic function $\text{Li}_{1+d/2}(\lambda_0^2 {\rm e}^{-2imt})$ results in an edge singularity in $P(W)$ located at $W=2nm$, with $n=1, 2, \ldots$. 
Therefore the lowest threshold of $P(W)$ is at  $W=2m$ and the associated edge singularity is determined by the lowest term of such a Fourier series, and --- as discussed further below --- it takes the form $\propto \vartheta(W-2m) (W-2m)^{d/2-1}$, where $\vartheta(x)$ is the Heaviside step function. 
This heuristic argument is supported by the following analysis, which allows a calculation of the expression of the probability $P(W)$ for $W < 4 m$ beyond the stationary phase approximation. First of all we observe that if the (inverse) Fourier transform $\tilde f(\omega)$ of a function $f(t)$ vanishes identically for $\omega<\omega_0$, then the Fourier transform of ${\rm e}^{f(t)}$  is given by $\delta(\omega) + \tilde f(\omega)$ for $\omega < 2 \omega_0$. In order to see this, one can formally expand ${\rm e}^{f(t)} = 1 + f(t) + f^2(t)/2 +\ldots $ and note that, by virtue of the convolution theorem, the $n$-th term of this expansion contributes to the Fourier transform of ${\rm e}^{f(t)}$ only for $\omega\ge n\omega_0$. Accordingly, for the specific case we are interested in, 
$P(W)$ for $0<W<4m$ is proportional to the inverse Fourier transform $\tilde{\varphi}(W)$ of $\varphi(t)$, which, in turn, is given only by the inverse Fourier transform of the first term of the series in Eq.~(\ref{series}), since all the other terms contribute to $\tilde{\varphi}(W)$ only  for $W> 4m$.
Accordingly, $P(W)$ for $0<W<4m$, i.e., the profile $P_{\rm es}$ of the lowest edge singularity, is given by 
\eqa{ P_{\rm es}(W) & = \frac{1}{2} L^d \mathcal{F}(m_0,m) \int \frac{\rmd t}{2\pi} \; {\rm e}^{iWt} \int_k \lambda_k^2 {\rm e}^{-2i\o_k t} \nn \\ 
& = \frac{1}{2} L^d \mathcal{F}(m_0,m) \int_k \lambda_k^2 \, \delta(W-2\omega_k)\nn\\
& = \frac{(mL)^d}{m} \mathcal{F}(m_0,m) \; \chi\(\frac{W}{2m} \, ; \, \frac{m_0}{m}\), \label{esth}}
with
\eqa{ \chi(x;\rho) & \equiv  \frac{\Omega_d}{4(2\pi)^d}  \vartheta(x-1) (x^2-1)^{d/2-1} \times \nn \\
& \qquad  \times x \(\frac{\sqrt{x^2+\rho^2-1}-x}{\sqrt{x^2+\rho^2-1}+x}\)^2 \label{chi}.}
Note that the first term of the expansion of the exponential contributes to $P(W)$  with ${\cal F}\delta(W)$, as expected.
For $x\simeq 1^+$, i.e., upon approaching the threshold $W \to 2 m^+$, Eq.~\reff{chi} becomes
\eq[chitresh]{
\begin{split}
\chi(x\to 1^+;\rho) = \frac{\Omega_d}{4(2\pi)^d}\lambda_0^2 \; [2(x-1)]^{d/2-1},
\end{split}}
[where $\lambda_0 = (\rho-1)/(\rho+1)$] which renders the algebraic behaviour of $P(W\to 2m)$ that we anticipated above and that can be alternatively inferred by taking the inverse Fourier transform of the function $\varphi(t)$ calculated within the stationary-phase approximation $m t \gg 1$ reported in Eqs.~\reff{asympt} and~\reff{asymptphi}.
In deriving these results, we assumed that the system size is large but still finite, otherwise the expansion of the exponential ${\rm e}^{- L^d \varphi(t)}$ is no longer justified and the problem can be studied only in terms of large deviations of the intensive work $w = W/L^d$, as discussed in Ref.~\onlinecite{GS12}.

The behaviour of $P(W)$ at the threshold is generically non-analytic: when the threshold is approached from above, $P(W)$ diverges for $d<2$, attains a finite value for $d=2$, while it vanishes for $d>2$. Figure~\ref{fig1b} shows (for a choice of $m_0$ and $m$) the profile of the lowest edge singularity in $d=1$, 2, and 3 as obtained numerically by a direct evaluation of the integral in Eq.~\reff{PWdisc}, with $\varphi$ and ${\cal F}$ given in Eq.~\reff{comp1} and \reff{fid}, respectively. This numerical result (solid line) is compared with the analytic expression $P_{\rm es}$ (dashed line) provided by Eqs.~\reff{esth} and \reff{chi}. The agreement between these curves is excellent for $W<4 m$, as expected.
On the other hand, one can easily verify that the asymptotic expression for $W\to 2m$ in Eq.~\reff{chitresh} --- determined by the behaviour of $\varphi(t)$ for $m t \gg 1$ --- does not provide an accurate approximation of $P_{\rm es}$ as soon as one moves away from the threshold, as it is clear by comparing Eqs.~\reff{chi} and \reff{chitresh}, which display a qualitatively different dependence on $x$ and differ already at the linear order in $x-1$. In order to obtain a better approximation of the actual dependence on $x$ of the scaling function $\chi(x;\rho)$ it is necessary to account for the behaviour of $\varphi(t)$ also for $m t\sim 1$. Further below we discuss how to extend the analysis of the scaling limit to generic values of $m t$, highlighting the emergence of a scaling function which is expected to fully describe it.
In particular, the connection with the critical Casimir effect that we establish below for the present non-interacting model is a special case of the more general correspondence discussed in Ref.~\onlinecite{Gambassi2011} --- and exemplified therein for quenches in the quantum Ising chain in a transverse field --- which allows one to determine the behaviour of the edge singularity in interacting cases.
Notice, however, that one generically expect that corrections to the scaling limit --- which depend, e.g., on the lattice spacing $a$ and on additional microscopic parameters --- become increasingly relevant for values of $W$ and $2 m$ which are large on the microscopic scale. In this case, the behaviour of the system and therefore the statistic of the work is no longer solely determined by the long-wavelength modes in terms of which universality typically emerges.

The discussion in the previous paragraph assumes that $G(t)$ for large $L$ is given by Eq.~\reff{qqpf}, which is the case only if the sums $L^{-d}\sum_{k_n}$ over the wave vectors $k_n \propto n/L$, $n \in {\mathbb Z}^d$ allowed in a system of finite extent $L$ (and suitable boundary conditions) can be replaced by the integrals $\int_k$ indicated on the r.h.s. of Eq.~\reff{qqpf}. In particular, the discreteness of the set of allowed momenta $k_n$ implies that the lowest excitation energies $2 \omega_{k_n}$ of each single harmonic oscillator form a discrete set and that, accordingly, $W$ takes values in this set (see, e.g., the second line of Eq.~\reff{esth} with the integral replaced by the sum). For a given gap $m$ of the post-quench Hamiltonian, the difference between the possible subsequent values of $W$ (related to the inverse density of states) decreases quickly as $W$ increases above the threshold $m$. However, the first allowed value above the threshold $2m$ is given by $2\omega_{k_1}$ with $k_1^2\propto 1/L^2$. In order for the spacing $\Delta W \equiv 2\omega_{k_1}-2m$ to be negligible compared to the scale of variation of $P(W)$, set by the distance $2m$ between two subsequent thresholds (see the scaling form in Eq.~\reff{esth}), one has to require $\Delta W \ll 2m$ and therefore $L \gg m^{-1}$, i.e., the size $L$ of the system has to be always larger than the inverse gap of the post-quench Hamiltonian or, equivalently, to the correlation length $\xi$ within the system. This condition is never fulfilled at the critical point $m=0$, which is characterized by the fact that $P(W)$ varies on a scale set by $\sim 1/L$ and therefore the discrete set of allowed values of $W$ cannot be replaced as a continuum one close to $W\simeq 0$. In addition, the consecutive thresholds identified above merge into a single one at $W=0$ and the argument which allows one to relate $P(W)$ to the inverse Fourier transform of $\varphi(t)$ does no longer apply because all the subsequent terms of the expansion of the exponential, which are higher-order powers of $\varphi(t)$, contribute to $P(W)$. For these reasons we consider below only the case in which the inverse gap $m^{-1}$ is large on the microscopic length scale set, e.g., by the lattice spacing $a$, such that the continuum limit is justified, but it is still significantly smaller than the size $L$ of the system.

Following the general discussion done after Eqs.~\reff{fe} and \reff{fe1} about the relation between the generating function of the work statistics at a quantum quench occurring close to a critical point and the critical Casimir effect, here we analyze this connection in more detail. In particular, we would like to compare the form taken by $\varphi(t)$ (see Eqs.~\reff{comp1} and \reff{qqpf}) in the scaling limit
with the universal Casimir contribution to the free energy of a
classical fluctuating  (quadratic) field (equivalent to a system of quantum non-interacting bosons) confined in a film of finite thickness and with suitable boundary conditions.
It is well-known that the universal behaviour of this system  upon approaching the critical point can be described in terms of a small number of effective boundary conditions (or, more precisely, in terms of the so-called \emph{surface} universality classes\cite{diehl97,Gambassi2011}), depending on some gross features of the quench under consideration (e.g., breaking of some symmetry etc.).
In the context of quenches in quantum systems\cite{calabrese_07}  it has been argued and then shown explicitly for non-interacting bosons that the proper effective boundary condition for the field is of Dirichlet type (see also section III.B in Ref.~\onlinecite{SC10} for further discussion), at least for sufficiently ``deep" quenches, i.e., with an initial mass $m_0$ which is large compared to all the other scales at play in the system (i.e., $m$ and possible microscopic scales).
As we will see below, our comparison also support this conclusion. 

\begin{figure}[h!t] 
\centering
\includegraphics[width=.95\columnwidth]{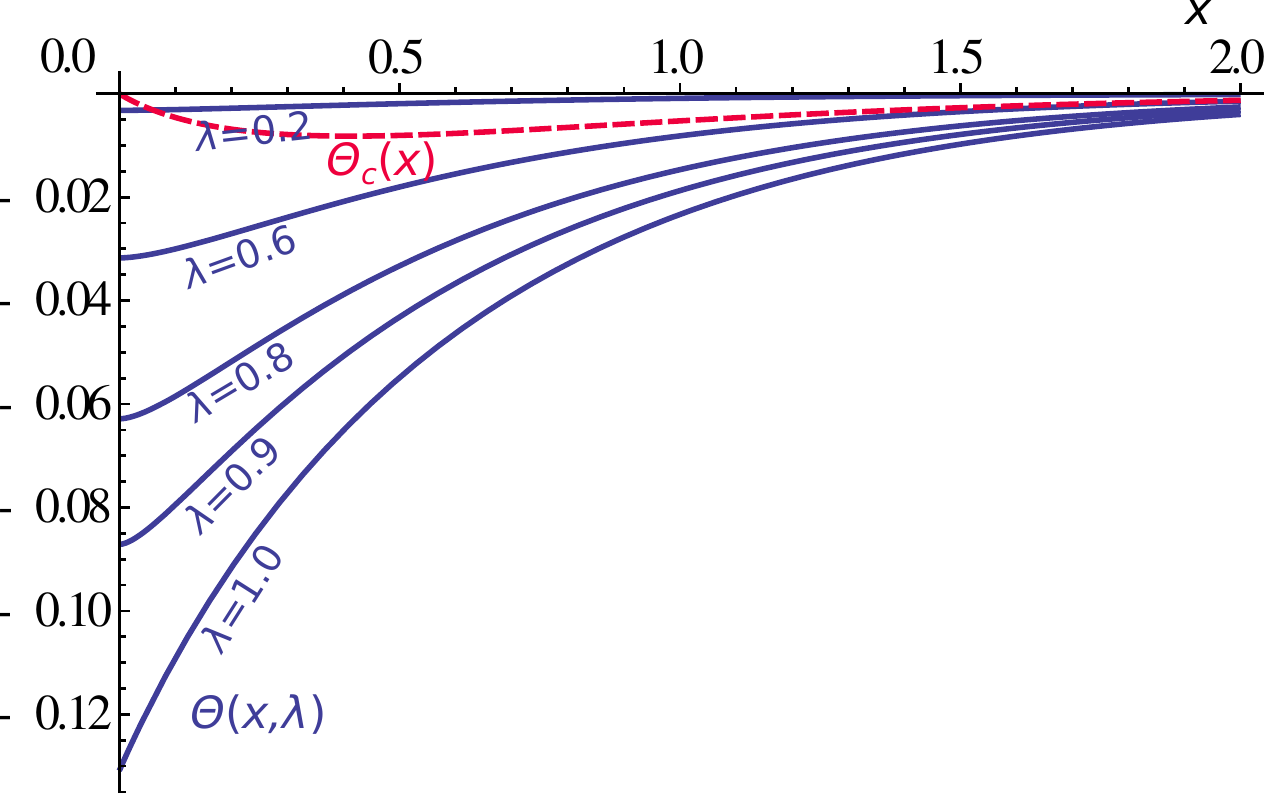}
\caption{\label{fig:theta} \small Scaling functions $\Theta(x;\lambda)$ (solid blue curves) for $\lambda=1$, 0.9, 0.8, 0.6, 0.2 and $\Theta_c(x)$ (dashed red curve) for the non-interacting bosonic model in one spatial dimension (see Eqs.~(\ref{Theta}) and (\ref{eq:vp-m0}) respectively). In contrast to $\Theta(x;\lambda)$, $\Theta_c(x)$ vanishes for $x=0$.}
\end{figure}

We first show that $\varphi(t)$ in Eq.~\reff{comp1} acquires a scaling behaviour for $t$ and $m$ large enough on a microscopic scale (but, contrary to the analysis leading to Eqs.~\reff{asympt}, \reff{asymptphi}, and \reff{chitresh}, with $mt$ assuming generic values).
In order for the integral in Eq.~(\ref{comp1}) to be well-defined also for $\lambda_k \to 1$, we will assume the prescription $t \to t - i 0^+$, which corresponds to introducing an ultraviolet cutoff in the momentum integrals.
Then, in this scaling limit, each integral in  Eq.~(\ref{series}) is determined by the low-$k$ behaviour ($k\ll a^{-1}$ where
$a$ the lattice spacing) of its integrand 
\eq[eq:eqSP]{\begin{split} 
&\int_k \lambda_k^{2n} {\rm e}^{-2ni\omega_k t} \\
& \quad \simeq \lambda_{0}^{2n} \frac{\Omega_d}{(2\pi)^d} \frac{2nit}{d} \int_{m}^\infty \rmd \omega \,(\omega^2-m^2)^{d/2} {\rm e}^{-2ni\omega t},
\end{split}}
where an integration by parts has been applied, in which the contribution for $\omega_k \to \infty$ vanishes due to the prescription $t = t - i 0^+$.
In Eq.~\reff{eq:eqSP}, $\lambda_k$ has been approximated by $\lambda_0$ on account of the fact that a direct inspection of its expression (see Eq.~\reff{eq:lambdak}) shows that $\lambda_k$ as a function of $\omega_k$ varies on a scale set by $m_0$, whereas the exponential factor varies on a scale $\omega\sim |t|^{-1}$. Accordingly, for $|t| \gg m_0^{-1}$ the variation of the former can be neglected.
Note that in principle one could account for the complete $k$-dependence of $\lambda_k$ in Eq.~\reff{eq:eqSP}, as it is done further below when discussing the case $m_0=0$; however, for the purpose of highlighting the connection with the critical Casimir effect it is more convenient to approximate first $\lambda_k \simeq \lambda_0$.
Now summing the series Eq.~(\ref{series}) and using Eq.~(\ref{eq:eqSP}), we obtain
\be
\varphi(t) = (it)^{-d} \Theta(imt;\lambda_0),
\label{eq:vp}
\ee
where
\eq[Theta]{\Theta(x;\lambda) \equiv -\frac{\Omega_d}{d(2\pi)^d} x^{d+1} \, \int_1^\infty\!\! \rmd s \frac{(s^2-1)^{d/2}}{\lambda^{-2}  {\rm e}^{2 x s}-1} .}
A plot $\Theta(x;\lambda)$ as a function of $x$ is provided in Fig.~\ref{fig:theta} for various values of $\lambda$ and $d=1$.  
The scaling form (\ref{eq:vp}) for $\varphi(t)$ can be directly compared with the known results for the critical Casimir effect in a film. 
Note that the model \reff{hamG} we are presently interested in involves a 
quadratic Hamiltonian of the real scalar field $\phi$ associated with the relative phase of the quasi-condensates and therefore its classical correspondent is a real scalar field, corresponding to the case $N=1$ of the $N$-component scalar field with $O(N)$ symmetry discussed in Ref.~\onlinecite{KD92}.
The universal Casimir contribution $f_C$ to the free energy per unit area $S=L^{D-1}$ and per temperature $\beta^{-1}$ of such a fluctuating classical  (quadratic) field in $D$ spatial dimensions confined in a film of thickness $R$ with Dirichlet boundary conditions (corresponding to the so-called ordinary surface universality class\cite{diehl86,diehl97}) is given by\cite{KD92}
\eq[eq:1]{ f_C(R,\xi;D) = R^{-(D-1)} \Theta_O(R/\xi), }
where $\xi$ is the correlation length of the fluctuations of this classical  field in the bulk (i.e., for $R\to \infty$) and
\eq[eq:2]{ \Theta_O(x) \equiv - \frac{x^D}{(4\pi)^{(D-1)/2} \Gamma\(\frac{D+1}{2}\)} \int_1^\infty\!\rmd s\, \frac{(s^2-1)^{\frac{D-1}{2}}}{ {\rm e}^{2 x s}-1} }
is the associated universal scaling function; see, e.g., Eq.~(6.6) of Ref.~\onlinecite{KD92}. 
By comparing Eqs.~(\ref{eq:vp}) and (\ref{Theta}) with Eqs.~(\ref{eq:1}) and (\ref{eq:2}) one concludes that
$\varphi(t)$ can be cast in the form
\eq[eq:3]{ \varphi(t) = f_C(R = it + 0^+, \xi = m^{-1};D=d+1), }
for $\lambda_0\to \pm 1$, i.e., either $m_0 \gg m$ or $m_0 \ll m$.
However, we shall see below that the latter case actually requires accounting for corrections to $\lambda_k$ which have been neglected in deriving Eq.~\reff{eq:eqSP}.
Equation~\reff{eq:3} demonstrates, for this specific case of the bosonic field, the connection\cite{Gambassi2011} between the generating function $G(t)$ and the critical Casimir effect free energy $f_C(R)$ of the corresponding classical model, which we discussed in section~\ref{sec:1}. 
As anticipated above, the scaling function $\Theta(x,\lambda)$ which controls the behaviour of $\varphi(t)$ (see Eqs.~\reff{eq:vp} and \reff{Theta}) coincides with the one $\Theta_O$ of the critical Casimir force in the presence of Dirichlet boundary conditions (see Eq.~\reff{eq:2}) only in the limit $m_0\gg m$ of a deep quench. In this case, the overall amplitude of the edge singularity, which is set by $\cal F$ according to Eq.~\reff{esth}, is exponentially suppressed as a function of $(m_0L)^d C_d$ (see Eq.~\reff{F:scal:asy}), where $C_d$ generically takes rather small values. Accordingly, in order for this feature to be observable one has to be able to detect somewhat rare events.
Furthermore, based on the discussion done after Eq.~\reff{asymptphi}, the behaviour of $P(W)$ for $2m < W < 4m$  is generically determined by the inverse Fourier transform of $-L^d \varphi(t)$ and therefore of the inverse Fourier transform of the associated universal scaling function $\Theta$ (see Eq.~\reff{eq:vp}). 
In particular, from Eqs.~\reff{eq:vp} and \reff{Theta} one finds that $P(W)$ takes the scaling form of Eq.~\reff{esth} with a scaling function $\chi'$ 
given by
\eq[chi:cas]{\chi'(x;\rho) = \frac{\Omega_d}{4 (2\pi)^d}\lambda_0^{2} x \left(x^2-1\right)^{d/2-1}}
for $1<x<2$ and $\lambda_0 = (\rho-1)/(\rho+1)$. Compared to the scaling function in Eq.~\reff{chitresh}, which accounts only for the behaviour of $\varphi(t)$ for $mt \gg 1$ and it is therefore accurate only for $x\to 1^+$, $\chi'$ provides a better approximation of the actual scaling function in Eq.~\reff{chi} and for $\rho \gg 1$ reproduces it in the full range $1<x<2$. 
Note that, according to this relation between $P(W)$ and $\Theta(x;\lambda)$,
the leading behaviour of $P(W)$ upon approaching the threshold, i.e., for $W \to 2 m^+$ depends on the behaviour of the scaling function $\Theta$ at large (imaginary) values of the scaling variable $mt$, which is connected to the asymptotic behaviour of $f_C(R)$ at large (real) values of $R$.
In fact, rather generally, the relation between $P(W)$ and $G(t)$ in Eq.~\reff{0a} implies that $G(t)$ as a function of the complex variable $t = t_R + i t_I\in {\mathbb C}$, with $t_{I,R}\in {\mathbb R}$, cannot have poles in the lower complex half plane $t_I < 0$ because $P(W<0)=0$ (here we assume that $W$ is referred to the threshold value $\Delta E_{gs}$). In addition, $G(t)$ has the symmetry $G^*(t)=G(-t^*)$. For the specific model we are presently interested in, $G(t)$ in a finite volume (e.g., assuming periodic boundary conditions) is the product over the allowed wave-vectors $k$ of the generating function $G_k$ of each single mode, i.e., $G(t) = \prod_{k} G_k(t)$, with $G_k(t)$ given in Eq.~\reff{qqptext}. Each of these factors is characterized by a \emph{branch cut} whenever the argument of the square root (or, equivalently, of the logarithm in $\ln G(t)\propto \varphi(t)$ in Eq.~\reff{qqpf} before taking the limit $L\to\infty$ which turns $\sum_k$ into $L^d \int_k $) becomes real and negative, i.e., for $t_I>(\ln \lambda_k^{-2})/(2 \omega_k)\ge 0$ and $t_R = t_R^{(n)} \equiv n \pi/(2 \omega_k)$, with $n\in {\mathbb Z}$.  (In passing, we mention that the analogous singularities for the quantum Ising chain in a transverse field are represented by simple \emph{poles}\cite{Heyl2012b}.) For a given mode $k$, these cuts are parallel to the imaginary axis, include it, and are displaced along the real axis. As expected, however, they are all located in the upper half plane $t_I > 0$. Accordingly, $G(t)$ and $\varphi(t)$ are analytic in the lower half plane $t_I < 0$ and therefore the behaviour of $\varphi(t)$ at large $t$ along the real axis (with a vanishingly small negative real part $-i0^+$) can be obtained by analytic continuation of the corresponding behaviour of the function along the negative imaginary axis $- i R$ with $R>0$, where it coincides with the critical Casimir force $f_C$ acting within a film of thickness $R$, as anticipated in Eq.~\reff{eq:3}.
%
%
%%%%%%%%%%%%%%%%%%%%%%%%%%%%%%%%%%%%%%%%%%%
%%%%%%%%%%%%%%%%%%%%%%%%%%%%%%%%%%%%%%%%%%%
%
%

%
The results discussed above are valid in the scaling limit and for $|t|\gg m_0^{-1}$, which allowed the approximation $\lambda_k\simeq \lambda_0$ in Eq.~\reff{eq:eqSP}. The leading correction to this approximation takes the form of a renormalization of $t$ and $\lambda_0$. In fact, the first-order correction to $\lambda_k$ as a function of $\o_k$ can be expressed in exponential form
\eq[eq:lambdacorr]{\lambda_k=\lambda_0 {\rm e}^{-2(\o_k-m)/m_0} [1+\mathcal{O}((\o_k-m)^2/m_0^2)]}
and therefore it translates  into an imaginary shift of $t$ and a rescaling of $\lambda_0$ in Eq.~(\ref{eq:vp}):
\eq[eq:effpar]{t\mapsto t_{\text{eff}}\equiv t - 2i/m_0 \quad \text{and} \quad \lambda_0 \mapsto \lambda_{\text{eff}} \equiv \lambda_0 {\rm e}^{2m/m_0},}
i.e., to the introduction of the effective parameters $t_{\text{eff}}$ and $\lambda_{\text{eff}}$. In particular, the emergence of $t_{\text{eff}}$ confirms explicitly the expectation\cite{calabrese_07} that a large but finite value of $m_0$ amounts at an extrapolation length\cite{GS12} 
\eq[eq:lext]{\ell_{\text{ext}} \equiv m_0^{-1}}
at each of the two surfaces of the film in the boundary formulation of the quantum quench problem; $\ell_{\text{ext}}$ accounts for the fact that the fixed-point boundary condition (in this case of Dirichlet type for $m_0$ large enough) is not effectively imposed at the boundaries of the film but on displaced effective boundaries. 
Analogously, the parameter $\lambda_{\text{eff}}$ accounts for the effective ``strength'' of the boundary condition and it approaches 1 
for $m_0 \gg m$, i.e, $\rho = m_0/m \gg 1$. 
The actual effect of $m_0$ taking a finite value  $\gtrsim m$ (i.e., $\rho \simeq 1$) on the behaviour of the probability distribution $P_{\rm es}(W)$ for $2m<W<4m$ is described by Eq.~\reff{esth} with the scaling function $\chi$ in Eq.~\reff{chi}, which was derived by considering the full $k$-dependence of $\lambda_k$ in Eq.~\reff{series}, while it is not entirely captured by the scaling function $\chi'$ in Eq.~\reff{chi:cas} which corresponds to $\ell_{\text{ext}}=0$.
However, part of this effect can be accounted for by taking the inverse Fourier transform of Eq.~\reff{eq:vp} with the effective parameters in Eq.~\reff{eq:effpar}. Because of the translation $t \mapsto t_{\rm eff} = t - 2 i \ell_{\rm ext}$, this transform acquires an overall multiplicative factor $\rme^{- 2W\ell_{\rm ext}}$ compared to the case with $\ell_{\rm ext} = 0$ while $\lambda_0 \mapsto \lambda_{\text{eff}} = \lambda_0 \rme^{2/\rho}$ according to Eq.~\reff{eq:effpar}. The resulting $P(W)$ takes the same scaling form as in Eq.~\reff{esth} with a scaling function $\chi''$ given by \eq[eq:cas:lext]{\begin{split} \chi''(x,\rho)  & =  \rme^{-4 x/\rho} \chi'(x;\rho)|_{\lambda_0 \mapsto \lambda_{\rm eff}} \\ 
& = \rme^{-4 (x-1)/\rho} \chi'(x;\rho).\end{split}} 
With this additional exponential factor, the difference between $\chi''$ and the actual scaling function $\chi$ turns out to be of order 
${\cal O}([(x-1)/\rho]^2)$, with a significant improvement compared to the scaling function $\chi'$.
In summary, while the knowledge of the asymptotic behaviour for large separations of the critical Casimir force in the classical system is sufficient to predict the leading behaviour of the probability $P(W)$ very close to its lower threshold, the knowledge of the full scaling function $\Theta$ provides the necessary information to predict $P(W)$ with a certain accuracy up to the second threshold, while additional improvements can be obtained by considering the emergence of an extrapolation length and of an effective surface coupling within the corresponding classical system.

Upon decreasing $m_0$, the extrapolation length $\ell_{\text{ext}}=m_0^{-1}$ which controls the corrections mentioned above grows, the influence of the corrections increases and eventually one can no longer  neglect the full $k$-dependence of $\lambda_k$, which is particularly strong for $m_0=0$. In this case, $\lambda_k(\omega_k)|_{m_0=0}$ has to be accounted for explicitly in Eq.~(\ref{eq:eqSP}) where it amounts to an additional multiplicative factor that is integrated by parts in the next step. As a result, one finds the same scaling behaviour as in Eq.~(\ref{eq:vp}), but with
a new scaling function given by
\eqa{\Theta_c(x) \equiv -\frac{\Omega_d}{d(2\pi)^d} x^{d} & \int_1^\infty \rmd s \frac{(s^2-1)^{d/2}}{(\sqrt{s^2-1}+s)^4  {\rm e}^{2 x s}-1}\times \nn \\ 
& \quad \times \(\frac{2}{\sqrt{s^2-1}}+x\) \label{eq:vp-m0} .}
Differently from the function $\Theta$ in Eq.~(\ref{Theta}), the present scaling function vanishes for $x\to 0$, i.e., if the post-quench Hamiltonian approaches the critical point, with 
\eq{\Theta_c(x\to0) =
\begin{cases}
- x^d \frac{\Gamma(1-\tfrac{d}{2})}{2 d (4\pi)^{d/2}} \left[\frac{2\Gamma(\tfrac{d+1}{2})}{\sqrt{\pi}\Gamma(1+\tfrac{d}{2})}-1\right]& \mbox{for}\ d<4,\\[4mm]
-x^4 \frac{\Gamma(d-4)}{2^d(4\pi)^{d/2}\Gamma(d/2)} & \mbox{for}\ d>4.
\end{cases}}
This behaviour is actually expected since for $m\to 0 = m_0$ the quench becomes vanishingly small and therefore no work is done on the system and the edge singularity disappears. Figure~\ref{fig:theta} shows plots of the universal scaling functions $\Theta(x;\lambda)$ and $\Theta_c(x)$ for $d=1$. Note, however, that the functional form of the behaviour of $\Theta_c$ for $x\gg 1$ is the same as the one of $\Theta(x;\lambda\neq0)$ and therefore the leading algebraic behaviour of the edge singularity as $W\to 2m$ in the case $m_0=0$ is the same as for $m_0\neq 0$, up to an overall amplitude, in agreement with Eq.~\reff{chitresh}. 
The inverse Fourier transform of the critical Casimir contribution $\varphi(t) = (it)^{-d}\Theta_c(imt)$ with $\Theta_c$ given in Eq.~\reff{eq:vp-m0} can be calculated and the probability distribution $P(W)$ for $2m < W < 4m$ acquires the scaling form in Eq.~\reff{esth} (with $m_0=0$), with the scaling function $\chi$ replaced by $\chi_c(x) = \chi(x,0)$. This scaling function for $x \to 1^+$  has the same behaviour as $\chi$, $\chi'$ and $\chi''$ up to an overall multiplicative factor, because the asymptotic behaviours of $\Theta(x;\lambda_0)$ and $\Theta_c(x)$ for $x \gg 1$ are the same. However, depending on the actual spatial dimensionality $d$ of the system, this common behaviour might emerge only very close to the threshold $x=1$ because of the presence of rather strong corrections $\chi_c(x)/\chi''(x,\rho) = \lambda_0^{-2}[1-\sqrt{8(x-1)} + {\cal O}(x-1)]$ which mirror those in the asymptotic behaviour of $\Theta_c(x\gg 1)$ compared to $\Theta(x\gg 1,\lambda_0)$ clearly visible in Fig.~\ref{fig:theta}. More pronounced differences emerge upon moving further away from the threshold, which are connected to the markedly different behaviours of the corresponding scaling functions for $x\lesssim 1$. 
Note that in the present case $m_0=0$, the overall amplitude ${\cal F}(m_0=0,m)$ of $P(W)$ (see Eq.~\reff{esth}) is exponentially decreasing as a function of $(mL)^dC_d$, as indicated by Eq.~\reff{F:scal:asy}, where $C_d$ generically takes rather small values and $mL$ has been assumed to be large compared to 1 in order for the present analysis to be valid. Accordingly, in order for the edge singularity $P(W)$ to be observable one has to be able to detect somewhat rare events.

\

\

\

\section{Thermal quantum quench}\label{sec:4}

In this section we extend our investigation of the statistics of the work to the case in which the initial state is not the ground state of the pre-quench Hamiltonian, but a thermal state at temperature $\beta^{-1}$. 
The analysis which follows is an extension  of the one presented in Ref.~\onlinecite{Lutz1008} for a generic quench of a single oscillator to the case of a collection of non-interacting harmonic oscillators (see Eq.~\reff{ham1}) subject to an instantaneous quench. Due to the absence of interaction,
the generating function $G_{\rm th}(t)$ of the thermal $P(W)$ is still given by the product $G_{\rm th}(t)=\prod_k G_{{\rm th},k}(t)$ of the generating functions for each single mode $k$, which is calculated in Eq.~(\ref{tqpf0}) of the appendix~\ref{app}. The result is
\begin{widetext}
\eqa{\frac{\ln G_{\rm th}(t)}{L^d} = & -if_b \; t + \int_k [\ln(1- {\rm e}^{-\beta \o_{0k}}) + \frac12 \ln(1-\lambda_k^2)] - \nn \\
& - \frac{1}{2} \int_k \ln\[(1- {\rm e}^{-2(\beta {-i t}) \o_{0k}}\lambda^2_k) - 2 (1-\lambda_k^2) {\rm e}^{-(\beta {-i t}) \o_{0k}-i\o_k t}  - (\lambda^2_k - {\rm e}^{-2(\beta {-i t}) \o_{0k}}) {\rm e}^{-2i\o_k t}\] \label{tqpf} . }
\end{widetext}
Consistently with the conventions introduced after Eq.~\reff{qqpf}, the term $-i f_b t$ on the r.h.s. above can be omitted if one measures $W$ starting from the difference $\Delta E_{gs}$ between the energies of the ground states of the post- and pre-quench Hamiltonians by renaming $W$ as $W + \Delta E_{gs}$.

We repeat the calculation of the cumulants of the distribution $P(W)$ for the thermal case and derive the leading corrections they induce to the ordinary quench results in the limit of small temperatures ($\beta\to\infty$). In particular, for the mean work $ \mu'_W = i G'_{\rm th}(0)$ one finds
\eq[eq:muP]{\mu'_W = \mu_W - L^d(m_0^2 - m^2) \int_k \frac{ {\rm e}^{-\beta \o_{0k}}}{2 \o_{0k} (1- {\rm e}^{-\beta \o_{0k}})}, }
where $\mu_W=i G'(0)$ is the work done during the corresponding ordinary quench ($\beta^{-1}=0$), given in Eq.~\reff{k1}. As expected, the difference $\mu'_W-\mu_W$ vanishes at low temperatures $\beta m_0 \gg 1$ as
\eq{ \mu'_W-\mu_W \simeq - (m_0L)^d\, \frac{m_0^2 - m^2}{2m_0} \,\frac{ {\rm e}^{-\beta m_0}}{(2\pi\beta m_0)^{d/2}}.}
Similarly, for the variance ${\sigma'}_W$ one finds
\eq{ {\sigma'}^2_W = \sigma^2_W + L^d(m_0^2 - m^2)^2 \int_k \frac{ {\rm e}^{-\beta \o_{0k}}}{2 \o^2_{0k} (1- {\rm e}^{-\beta \o_{0k}})^2}, }
where $\sigma^2_W$  is the variance in the corresponding  ordinary quench, given in Eq.~\reff{k2};  The lowest-order correction at low temperatures is
\eq{ {\sigma'}^2_W - \sigma^2_W \simeq (m_0 L)^d \, \frac{(m_0^2 - m^2)^2}{2m_0^2} \frac{ {\rm e}^{-\beta m_0}}{(2\pi\beta m_0)^{d/2}}.}
These expressions show that thermal corrections are generically of order $\sim {\rm e}^{-\beta m_0}$. 

As in the case of the ordinary quench, it is interesting to compare the work done during the thermal quench with the work $W_{\rm rev}$ required in order to change $m$ from its pre- to its post-quench value in a reversible way, i.e., while keeping equilibrium with the bath at temperature $\beta^{-1}$. 
According to thermodynamics, the latter is given by the change 
\eq[eq:DF]{\Delta F = F_\beta(g)-F_\beta(g_0)}
of the free energy $F_\beta(g)$ of the system, where we indicate generically by $g$ the parameter of the quench. In the case we are presently interested in we identify $g$ with $m^2$ and $g_0=m_0^2$, while $F_\beta(g)$ can be expressed, for large $L$, as
\eq[eq:Fho]{\begin{split}
F_\beta(g) =& -\beta^{-1} \ln {\rm Tr}\{\rme^{-\beta H(g)}\}\\
=& - \beta^{-1} L^d \int_k \ln Z(\omega_k(m=\sqrt{g})),
\end{split}}
where $Z(\omega) = \sum_{n=1}^\infty \rme^{-\beta\omega(n+1/2)} = \left[2 \sinh(\beta\omega/2)\right]^{-1}$ is the partition function of a single harmonic oscillator with characteristic frequency $\omega$. The reversible work is therefore given by
\eq[eq:Wrev]{W_{\rm rev} =  \Delta F = \beta^{-1} L^d \int_k \ln\frac{\sinh[\beta\omega_k/2]}{\sinh[\beta\omega_{0k}/2]},}
(where $\o_k \equiv \omega_k(m=\sqrt{g})$ and $\o_{0k} \equiv \omega_k(m_0=\sqrt{g_0})$) and it has to be compared with $\langle W \rangle = \mu'_W$, which renders $\Delta E_{gs}$ reported after Eq.~\reff{qqpf} for $\beta\to\infty$. Taking into account Eqs.~\reff{eq:muP}, \reff{k1}, and the shift of $W$ by $\Delta E_{gs}$ (such that $\la W \ra = \mu'_W + \Delta E_{gs}$), one eventually finds
\eq[eq:Wth]{\langle W \rangle = \frac{g-g_0}{4}  L^d \int_k \frac{1}{\o_{0k}}\frac{1}{\tanh(\beta \o_{0k}/2)}.}
Interestingly enough, these expressions for $\langle W\rangle$ and $W_{\rm rev}$ are similar to those for the thermal quench of the Ising chain discussed in Ref.~\onlinecite{Dorner2012}. In addition, Eq.~\reff{eq:Wth} shows that the dependence of $\langle W\rangle$ on $g=m^2$ is linear, as in the case of the ordinary quench (see section~\ref{sec:2}).
In fact, according to the argument outlined after Eq.~\reff{k1}, 
\eq[eq:Ww]{\begin{split}
 \la W \ra &= \frac{{\rm Tr}\{ \rme^{-\beta H_0}(H - H_0) \}}{{\rm Tr}\{\rme^{-\beta H_0}\}} \\
&= (g-g_0)\frac{{\rm Tr}\{ \rme^{-\beta H_0}\partial H_0/\partial g_0\}}{{\rm Tr}\{\rme^{-\beta H_0}\}}  \\
&= (g-g_0) F'_\beta(g_0),
\end{split}}
and therefore the non-equilibrium work can actually be expressed in terms of the equilibrium free energy $F_\beta(g_0)$. With the proper identifications, the same relation holds for the case of the Ising model studied in Ref.~\onlinecite{Dorner2012} (where the strength of the transverse field plays the same role as $g=m^2$ here).
Accordingly, the irreversible work $W_{\rm irr}$\cite{A} can be expressed as
\eq[eq:Wirr2]{
\begin{split}
W_{\rm irr} & = \langle W \rangle - \Delta F \\
& =(g-g_0) F'_\beta(g_0) + F_\beta(g_0) - F_\beta(g) \ge 0,
\end{split}}
where the last inequality --- which is actually expected on the basis of thermodynamics --- follows from the fact that $F''_\beta(g) \le 0$, i.e., $F_\beta(g)$ is a concave function of $g$, as it can be verified from Eq.~\reff{eq:Fho}.

The irreversible work $W_{\rm irr}$ is responsible for an irreversible increase of entropy $\Delta S_{\rm irr}(g,g_0) = \beta W_{\rm irr}\ge 0$. 
For a small quench with $g=g_0+\delta g$, Eq.~\reff{eq:Wirr2} yields 
\eq[eq:DSirr]{\Delta S_{\rm irr}(g_0+\delta g,g_0) = - (\delta g)^2 \beta F''_\beta(g_0)/2 + {\cal O}(\beta (\delta g)^3),} 
in which the only contribution on the r.h.s. comes from $W_{\rm rev} = \Delta F$ because $\la W \ra$ depends linearly on $\delta g$. For a fixed (small) value of $\delta g$, the irreversible entropy production as a function of the pre-quench value $g_0$ (or, alternatively, of the post-quench one $g$) has been shown to display a pronounced maximum upon approaching the critical point (located at $g_0 = g_0^*$) in a thermal quench of the quantum Ising chain\cite{Dorner2012}. 
This behavior has been traced back to the fact that, as the energy gap closes at criticality, most of the work done during the quench is dissipated in exciting the system and this dissipation is responsible for the emergence of an intrinsic irreversibility\cite{Dorner2012} quantified by $\Delta S_{\rm irr}$.
Rather generally, we can infer from Eq.~\reff{eq:DSirr} that the emergence of such a maximum in $\Delta S_{\rm irr}$ is however related to the behavior of $F''_\beta(g_0)$ for $g_0\to g_0^*$ and therefore it is \emph{de facto} determined by the \emph{equilibrium} free energy $F_\beta(g_0)$ of the system at finite temperature $\beta$ upon approaching the (quantum) critical point $g_0 \to g_0^*$. Both in the present case of the bosonic Hamiltonian in Eq.~\reff{ham1} and for the Ising chain discussed in Ref.~\onlinecite{Dorner2012}, the coupling constant $g$  (and therefore $g_0$) couples the most relevant operator which drives the system away from the critical point and therefore $g$ plays the role that the deviation of the temperature $T$ from its critical value $T_c$ has for a classical phase transition at finite temperature. Accordingly,  the possible non-analytic behavior in $F''_\beta(g_0)$ is characterized by the critical exponent $\alpha$ of the specific heat $C \propto - F''_\beta(g_0) $, i.e., $C \sim |g_0-g_0^*|^{-\alpha}$ and therefore $\Delta S_{\rm irr} \sim \beta (\delta g)^2|g_0-g_0^*|^{-\alpha}$ at least as long as $|\delta g| \ll |g_0-g_0^*|$, such that higher-orders in the expansion in Eq.~\reff{eq:DSirr} are actually negligible.
For the Ising chain at zero temperature $\beta^{-1}=0$ --- which displays the same critical behavior as the 2$d$ classical Ising model at finite temperature --- $\alpha=0$ and the singularity is actually logarithmic. However, at finite temperature, there is no phase transition in the model and therefore the pronounced maximum which might be observed in $\Delta S_{\rm irr}$ as $\beta^{-1}\to 0$ is weakened, as it is clearly visible in Fig.~1 of Ref.~\onlinecite{Dorner2012}. 

For the system with Hamiltonian~\reff{ham1} we are presently interested in, we recall that the critical point is located at $g_0^*=0$ and $g_0$ as well as $g$ can take only positive values in order for the Hamiltonian to have a ground state.
The corresponding free energy $F_\beta(g)$ in Eq.~\reff{eq:Fho} and in $d$ spatial dimensions can also be expressed (after a suitable regularization of the UV behavior) in terms of the partition function 
$Z_{\rm cl} \propto \int {\cal D}\phi\,\exp\{-S[\phi]\}$
of a classical (real) field $\phi(x,\tau)$ with  action $S = \int_0^\beta\rmd \tau\int\rmd^d x \,{\cal L}$, where the (euclidean) lagrangian density ${\cal L}$ is given by Eq.~\reff{hamG} after the replacement $\Pi \mapsto \partial_\tau \phi(x,\tau)$ and $m^2\mapsto g$. The additional ``spatial" dimension $\tau$ is of finite extent $\beta$ and the field  $\phi$ satisfies periodic boundary conditions $\phi(x,\tau=0)=\phi(x,\tau=\beta)$ along the $\tau$-direction, i.e., at the boundaries of this effective film of thickness $\beta$. Note that this construction is analogous to the one discussed after Eq.~\reff{0} and therefore also in this case the free energy $F_\beta(g)$ for large $L$ decomposes as in Eq.~\reff{fe}: 
\eq[eq:Cper]{\beta F_\beta(g) = L^d [\beta \varepsilon_b + f^{(per)}_C(\beta)],} 
where $\varepsilon_b(g) = \int_k \omega_k/2$ is the energy density of the ground state, whereas $f^{(per)}_C$ (which depends, inter alia, on $g$) represents the finite-size contribution which vanishes for $\beta\to\infty$ and is responsible, close to the critical point, for the emergence of the universal critical Casimir effect. Compared with Eq.~\reff{fe} the ``surface" contribution corresponding to $2 f_s$ is absent here because the periodic boundary conditions do not break translational invariance across the ``film"\cite{fss1}.  Upon approaching the critical point $g_0 \to g_0^*$  one expects $f^{(per)}_C(\beta)$ to take the scaling form in Eq.~\reff{eq:1} where $D=d+1$, $\xi = m^{-1} = g^{-1/2}$, $R$ is replaced by $\beta$ and the scaling function $\Theta_O(x)$ (see Eq.~\reff{eq:2}) by the analogous function $\Theta_{per}(x)$ appropriate to periodic boundary conditions.   
$\Theta_{per}(x)$ is given by (see, e.g., Eq.~(6.8) in Ref.~\onlinecite{KD92})
\eq[eq:2bis]{\Theta_{per}(x) = - \frac{x^D}{(4\pi)^{(D-1)/2}\Gamma(\frac{D+1}{2})}\int_1^\infty\!\!\rmd s\frac{(s^2-1)^{\frac{D-1}{2}}}{\rme^{xs}-1},} 
as one can verify by analyzing Eq.~\reff{eq:Fho} after the subtraction of the bulk contribution $L^d\beta\varepsilon_b$ (see below Eq.~\reff{eq:Cper}).

Note that the specific heat $C\propto - F''_\beta(g)$ can be consequently decomposed as in Eq.~\reff{eq:Cper}. In particular,  upon decreasing the temperature, the effective thickness $\beta$ of the film diverges and only the bulk term contributes to $F''_\infty(g)$. Being the system effectively $d+1$-dimensional, one expects $C$ (and therefore $F''_\infty(g) = L^d \varepsilon''_b(g)$) to be characterized by a critical exponent $\alpha = \alpha_{d+1}$, where $\alpha_D \equiv 2 - D/2$ is the specific-heat critical exponent of a  (scalar and real) field in $D$ space dimensions. 
Accordingly, the expansion of $\varepsilon_b(g)$ for small $g$ contains  a leading non-analytic contribution $\sim g^{2-\alpha_{d+1}}$ in additional to regular, analytic terms.  
At a finite temperature $\beta^{-1}\neq 0$, instead, the $d+1$-dimensional system has a finite thickness: when the correlation length $\xi=g^{-1/2}$ of the field fluctuations is much larger than the thickness $\beta$ of the film --- which occurs sufficiently close to the critical point --- the system displays the critical behavior corresponding to $d$ spatial dimensions, with $\alpha = \alpha_d$ and therefore one expects $F''_\beta(g) \sim g^{-\alpha_d}$.
In fact, from a suitable expansion for small argument of $\Theta_{per}(x)$ in Eq.~\reff{eq:2bis} one can verify that the resulting expansion of  $f^{(per)}_C(\beta)$ in $g$ for $\beta^2 g \ll 1$ contains both $\beta$-dependent analytic contributions $\beta^{-d} [a_0 + a_1 \beta^2 g + a_2 (\beta^2g)^2 + \ldots]$  and a leading $\beta$-independent non-analytic contribution $\propto g^{d/2}[1 + {\cal O}(\beta\sqrt{g})]$. When inserted into Eq.~\reff{eq:Cper}, the latter term renders the expected singularity $\sim g^{-\alpha_d}$ of $F''_\beta(g)$  as $g\to 0$, which indeed prevails on the non-analytic behavior of $\varepsilon''_b$ discussed above.  The presence of these non-analytic contributions follows from general properties of the scaling function $\Theta_{O,per}(x)$ of critical Casimir forces\cite{KD92} and can be also verified  via a direct calculation of $F_\beta''(g)$ based on Eq.~\reff{eq:Fho}. 

Accordingly, the behavior that the \emph{irreversible} entropy production $\Delta S_{\rm irr}(g_0+\delta g,g_0)$ displays for small $\delta g$ upon varying $g_0$ and therefore the singularity developed upon approaching its critical point $g_0 = g_0^* = 0$, is completely controlled by \emph{equilibrium} critical exponents and a dimensional crossover\cite{fss1,fss2} --- resulting from the competition between the bulk and the finite-size term in Eq.~\reff{eq:Cper}  --- is observed by lowering the temperature $\beta^{-1}$, which is accompanied by  an increase of $\Delta S_{\rm irr} \propto \beta$.

\begin{figure}%[htbp]
\centering
\includegraphics[width=\columnwidth]{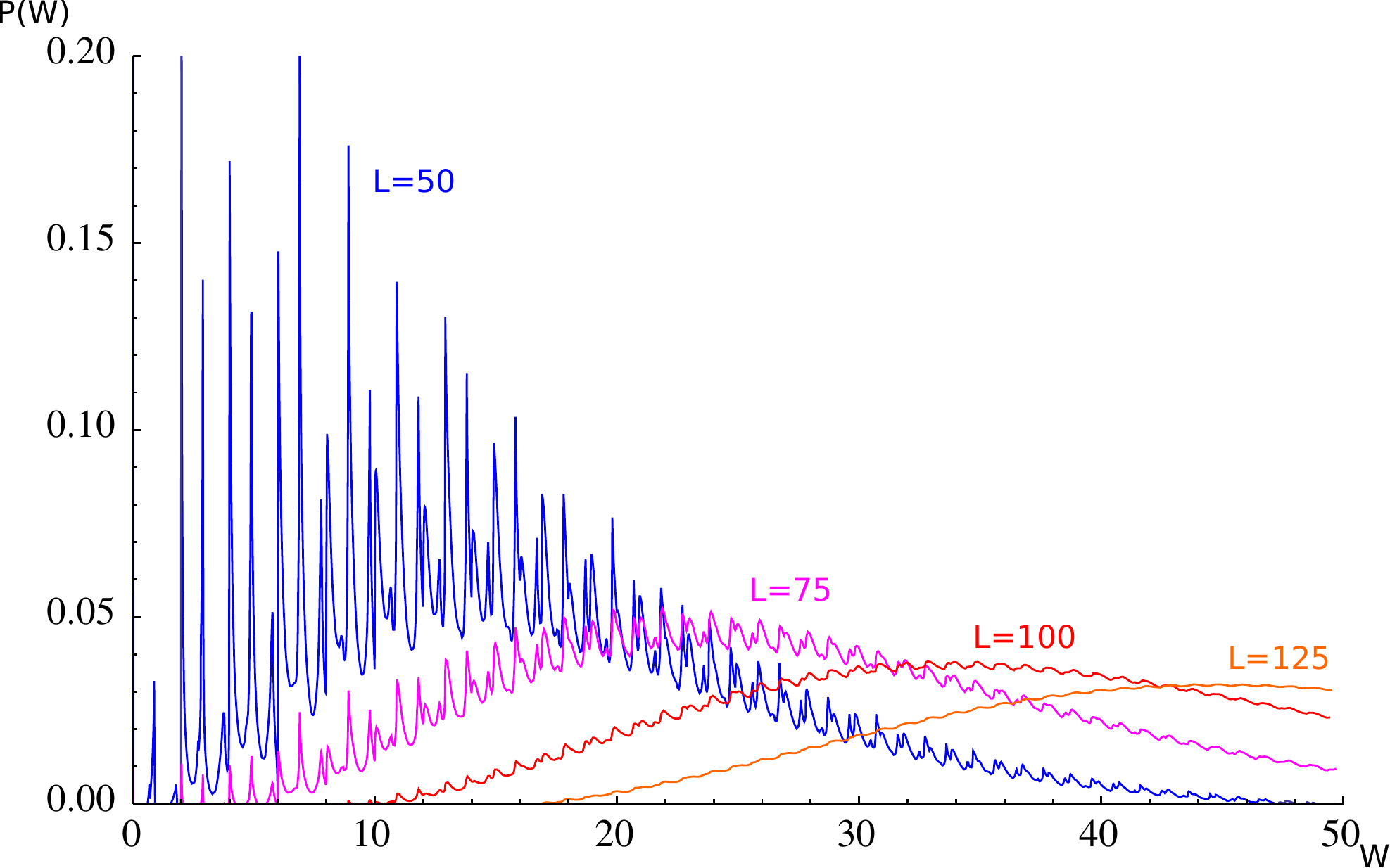}
\caption{\label{f2} \small 
Probability density $P(W)$ of the work for a thermal quantum quench in $1d$ with $m=1$, $m_0=0.1$, $\beta^{-1}=0.1$, and increasing system size $L=50$, 75, 100, and 125. Except for the non-zero temperature, all other parameters of the quench are the same as in Fig.~\ref{fig1}, with which the present figure should be compared. Note that, in contrast to Fig.~\ref{fig1}, there are peaks also for $W<2m$ which actually extend into the region $W<0$ (not shown here because these additional peaks are not visible on the vertical scale of the present plot due to their tiny height). All these peaks are of purely thermal origin and correspond to transitions originating from the thermal excitations of the initial state.}
\end{figure}

Figure~\ref{f2} shows $P(W)$ for a specific thermal quench of the 1$d$ model in Eq.~\reff{hamG} and for various values of $L$.  Differently from the zero-temperature case reported in 
Fig.~\ref{fig1} for the same values of the quench parameters but with $\beta^{-1}=0$, the lower threshold  at $W=2m$ has disappeared along with the regular structure of an equally spaced sequence of peaks. In addition, $W$ can now assume arbitrarily large negative values (though with an exponentially small probability) due to the fact that the system can be in an arbitrarily highly excited state of the pre-quench Hamiltonian, which can provide ($W<0$), rather than absorb ($W>0$), the energy required for the formation of post-quench excitations. Note that these features of Fig.~\ref{f2} are actually generic and their occurrence does not depend on the specific choice of the parameters of the quench as long as $\beta^{-1}\neq 0$.

According to the general discussion done in Ref.~\onlinecite{TH07}, we expect the distribution function $P(W)$ generated by $G_{\rm th}$ in Eq.~\reff{tqpf} to satisfy the fluctuation relations which characterize the statistics of the work done on isolated quantum systems. Indeed, in the present case of a quench, it is rather straightforward to show that the Crooks and Jarzynski fluctuation relations\cite{A} are satisfied, following essentially the analysis of Ref.~\onlinecite{Campisi2011}.  
$G(t)$ in Eq.~(\ref{1a}) can be cast in the form
\eq[n1]{G_{H_0\to H}(\beta;t) = \langle \rme^{-i W t}\rangle = \frac{\text{Tr}\{ {\rm e}^{-iHt} {\rm e}^{iH_0(t+i\beta)}\}}{Z_{H_0}(\beta)}}
where $Z_{H_0}(\beta) = \text{Tr}\{ {\rm e}^{-\beta H_0}\} $ is the partition function of the system with Hamiltonian $H_0$ at inverse temperature $\beta$. In this expression we have clearly indicated with a subscript that $G$ refers to the quench from $H_0$ to $H$. The cyclic property of the trace, implies 
\eq[n2]{ Z_{H_0}(\beta) G_{H_0\to H}(\beta;-t-i\beta) = Z_{H}(\beta) G_{H\to H_0}(\beta;t) , }
which, as expected, connects the generating function $G_{H_0\to H} $ of the work done by the ``forward'' quench to the one $G_{H\to H_0} $ of the ``backward'' quench, both originating from a thermal state of the corresponding pre-quench Hamiltonian (i.e., $H_0$ and $H$ respectively). Upon inverting the Fourier transform which defines the generating function, Eq.~(\ref{n2}) yields 
\eq[n3]{ \frac{P_{H_0\to H}(W)}{P_{H\to H_0}(-W)} = \frac{Z_H(\beta)}{Z_{H_0}(\beta)}{\rm e}^{\beta W} ,}
which expresses the Crooks relation for the present case. 
Analogously, Eq.~(\ref{n2}) at $t=0$ yields the Jarzynski relation\cite{A} $\la {\rm e}^{-\beta W} \ra = {\rm e}^{-\beta (F_H-F_{H_0})}$, where $F_H = - \beta^{-1} \ln Z_H(\beta)$ represents the free energy of the system with Hamiltonian $H$ in equilibrium at an inverse temperature $\beta$.

\section{Conclusions}\label{sec:5}

In this paper we discussed the statistics of the work done by a quantum quench of the gap (mass) of a system of non-interacting bosons, which capture the low-energy properties of a variety of experimentally relevant cases. We derived the expressions for the work probability distribution, focusing on the universal features which emerge close to the edge singularity at low values of the work and upon approaching the critical point of the post-quench Hamiltonian. We also studied thermal effects due to a non-zero initial temperature. 

For a system of non-interacting bosons, the calculation presented here first of all confirms the applicability of the analytical continuation from imaginary to real time of the predictions derived in the context of boundary field theory\cite{Gambassi2011}. In fact, an analogous analytic continuation has been shown to require some care in the case of the Ising chain in a transverse field quenched across its critical point\cite{Heyl2012b}.
By taking advantage of the relationship with the statistical properties of a classical  field fluctuating in a film geometry in $d+1$ space dimensions, we demonstrated that universal features emerge after a quantum quench of non-interacting bosons in $d$-dimensions upon approaching its critical (gapless) point, which are in fact linked to the critical Casimir effect characterizing the corresponding classical system. 
In order for the features discussed here to emerge, the size $L$ of the system has to be large compared to the correlation length $\xi$ of the post-quench Hamiltonian and both of them have to be large compared to the microscopic length scale $a$ (e.g., a possible lattice spacing) which characterizes the system.
From the experimental point of view, measurements of the work statistics after a quantum quench can provide insight into the excitation spectrum of the quantum system under probe. Our analysis shows that in low-dimensional systems that can be described (either exactly or approximately) by non-interacting, massive bosonic excitations, the work distribution is characterized by a pattern of ``fringes'' which can be easily identified. As a function of the work $W$, subsequent fringes are typically spaced by $2 m$, being $m$ the post-quench gap, while their overall ``amplitude" is set by the square ${\cal F}$ of the fidelity (see Eq.~\reff{fid}), which is exponentially suppressed as $L$ increases. Correspondingly, the extensive mean value $\langle W \rangle$ of the work grows as $\propto L^d$, giving rise to a broad  peak which carries most of the spectral weight\cite{GS12} and within which the various fringes are decreasingly visible (see Fig.~\ref{fig1}). 
Accordingly, in order to detect the features we have highlighted here it is necessary to investigate systems which are large on a microscopic scale but still of finite size $L \gg \xi\gg a$.  Depending on the choice of the parameters of the quench $m$ and $m_0$, which determine the $L$-dependence of the overall amplitude ${\cal F}$ (see Eq.~\reff{fid} and Fig.~\ref{fidel}), it is possible to find an optimal combination of their values which clearly allows the detection of the edge singularity, as shown in Figs.~\ref{fig1} and \ref{fig1b}.

In addition, as the space dimensionality $d$ of the model increases, the edge singularity associated to the lowest threshold at $W = 2m$ becomes increasingly milder and, upon approaching the threshold from above, it actually turns from a divergence for $d<2$ to a vanishing distribution for $d>2$  (see Fig.~\ref{fig1b}), with universal features encoded, inter alia, in the algebraic law which rules such an approach. In particular, while in one spatial dimension each peak is clearly distinguishable for the next one, as it decays to zero on a scale comparable to their separation, this is no longer the case in higher dimensions, where the weaker singularity results in a merging of the consecutive fringes.
Note that, in contrast to non-interacting fermionic systems where excited states cannot be more than singly or doubly occupied, in non-interacting bosonic systems multiple occupation of excited states leads to multiple excitation peaks. 
Accordingly, the work done on the system can in principle assume arbitrarily large values, though with a small probability, and therefore the support of the work probability distribution is not bounded (see the discussion in Ref.~\onlinecite{GS12}). 
In the case of a quench occurring from an initial thermal state, all the features highlighted above are lost and the probability distribution $P(W)$ of the extensive work $W$ takes the much less regular form reported in Fig.~\ref{f2} for a certain choice of the parameters of the quench. Differently from the ordinary quench, there is no lower threshold for the work, which can assume arbitrarily large values, though with small probability. Accordingly, no universal features are expected to emerge at sufficiently small $|W|$. 
By comparing the mean work done during the quantum quench with the work that would be done in equilibrium during a reversible variation of the parameter of the Hamiltonian from its pre- to its post-quench value, one can study the associated irreversible entropy production $\Delta S_{\rm irr}$, which is expected to increase significantly upon approaching the critical point\cite{Dorner2012}. Rather generally it turns out that the pronounced maximum showed by $\Delta S_{\rm irr}$ for narrow quenches and as a function of the quench parameter can be actually characterized in terms of the \emph{equilibrium} specific-heat exponent and that a dimensional crossover can be observed upon varying the temperature of the initial state.
An analysis along the one of Ref.~\onlinecite{GS12} of the properties of large fluctuations around the broad  distribution which develops upon increasing $L$  is left for future investigations.

The possibility to  measure experimentally the work statistics would enhance our understanding of the excitation spectrum after a quantum quench and would provide an intriguing check of universality which is the cornerstone of the proposed connection between classical equilibrium physics in film geometries (and the associated critical Casimir effect) and non-equilibrium processes exemplified by the quantum quench problem. 
Even though no experimentally viable technique is presently available for measuring the statistic $P(W)$ for many-body systems suitable extensions of recent proposals\cite{Dorner2013,Mazzola2013} might provide experimental access to its generating function $G(t)$.

\acknowledgments

S.S. acknowledges financial support by SISSA -- International School for Advanced Studies under the ``Young SISSA Scientists' Research Projects'' scheme 2011-2012 and by the ERC under Starting Grant 279391 EDEQS. 
A.G. is supported by MIUR within ``Incentivazione alla mobilit\`a di studiosi stranieri e italiani residenti all' estero''.  A.G. and A.S. are grateful to KITP for hospitality. This research was supported in part by the National Science Foundation under Grant No. PHY11-25915.
A.G. is grateful to J. Goold for useful correspondence.

\appendix
\section{Derivation of $G(t)$ for a single harmonic oscillator.}\label{app}

In this appendix, we consider a single harmonic oscillator described by the Hamiltonian
\eq[ham0]{H = \frac{1}{2} p^{2} + \frac{1}{2} \omega^{2} q^{2},}
in terms of the momentum $p$ and the position $q$ operators and we calculate the characteristic function $G(t)$ of the work done by a quantum quench of its frequency from $\o_0$ to $\o$. The initial state is considered to be either the ground-state $|\Psi_0\ra$ corresponding to frequency $\o_0$ or a thermal state at inverse temperature $\beta$, always at frequency $\o_0$. This problem is a special case of the one thoroughly addressed in Ref.~\onlinecite{Lutz1008} where the frequency is assumed to be any function of time $\o(\tau)$. Here, exploiting the simple step-like dependence of $\o$ on the time in the quantum quench protocol, we will first present a calculation of $G(t)$ for the case in which the initial state is the ground-state $|\Psi_0\ra$ by expanding the latter in the basis of eigenstates of the post-quench Hamiltonian $H$. Then we will repeat this calculation by using the coherent state formalism which has the advantage of being elegant, significantly simpler and straightforwardly generalizable to the 
thermal case, which we consider at the end of this appendix.

Let us start with the calculation of $G(t)$ 
by using the expansion of the initial state $|\Psi_0\ra$ in Eq.~(\ref{0}) on the eigenbasis $\{|n\ra\}$ of the post-quench Hamiltonian:
\eq[a0]{G(t)= {\rm e}^{i\o_0 t/2}\sum_{n=0}^\infty {\rm e}^{-\(n+\frac{1}{2}\) i \o t} |\la n|\Psi_0\ra|^2 ,}
where we used the fact that for a single harmonic oscillator $H|n\ra = (n+1/2)\o|n\ra$ and $E_0 = \o_0/2$. The first step is the calculation of the amplitudes $\la n |\Psi_0\ra$: since both the pre- and post-quench Hamiltonians are quadratic in the position and momentum operators $q$ and $p$, the relation between pre- and post-quench creation-annihilation operators $a_0,\,a_0^\dg$ and $a,\,a^\dg$ (which are linear combinations of $q$ and $p$) is linear. More specifically it is given by the single-boson version of the well-known Bogoliubov transformation
\eq[Bog0]{a_0 = \mu a + \nu a^\dg  \quad \text{ with } \quad |\mu|^2 - |\nu|^2 = 1 ,}
where the condition on $\mu$ and $\nu$ is due to the requirement of having canonical commutation relations for the two pairs of creation-annihilation operators. The Bogoliubov coefficients $\mu,\nu$ for the quantum quench problem are real and given by\cite{calabrese_07}
\eq[bog]{\mu = \frac{1}{2} \(\sqrt{\frac{\o_0}{\o}} + \sqrt{\frac{\o}{\o_0}}\) , \quad 
\nu = \frac{1}{2} \(\sqrt{\frac{\o_0}{\o}} - \sqrt{\frac{\o}{\o_0}}\).}

The amplitudes $\la n | \Psi_0\ra$ can be obtained readily by solving the recurrence relation they satisfy. The latter 
can be derived by taking into account that $a_0|\Psi_0\rangle = 0$ and $|n\rangle = (a^\dg)^n |0\rangle/\sqrt{n!}$, which yield:
\eqa{&\la n | \Psi_0\ra 
=\frac{1}{\sqrt{n!}} \la 0|a^{n-1}(\mu a_0-\nu a_0^\dg)| \Psi_0\ra \nn\\ 
&=-\frac{\nu}{\sqrt{n!}} \la 0|(n-1) \mu a^{n-2}+a_0^\dg a^{n-1}|\Psi_0\ra \nn\\
&=-\mu\nu \sqrt{\frac{n-1}{n}} \la n-2|\Psi_0\ra-\frac{\nu}{\sqrt{n!}} \la 0 |(\mu a^\dagger +\nu a)a^{n-1}| \Psi_0\ra \nn\\
&=-\mu\nu\sqrt{\frac{n-1}{n}} \la n-2 |\Psi_0\ra - \nu^2 \la n|\Psi_0\rangle,}
where we made use of the relation Eq.~(\ref{Bog0}) between $a$ and $a_0$ which implies the commutation relations $[a,a_0^\dagger]=\mu$ and $[a^n,a_0^\dg] = n\mu a^{n-1}$. We thus find the recurrence relation
\eq[recr]{\la n | \Psi_0\ra = -\lambda \sqrt{\frac{n-1}{n}} \la n-2|\Psi_0\ra ,}
with
\eq{\lambda \equiv \frac{\nu}{\mu} = \frac{\o_0-\o}{\o_0+\o} ,}
which allows us to determine the sequence $ \la n | \Psi_0\ra $ from its first two elements. A direct calculation based on the fact that the ground-state wave function $\Psi_{\rm GS}(q)$ of the harmonic oscillator with frequency $\o$ is $ \Psi_{\rm GS}(q) \propto \exp(-\o q^2/2)$, yields $\la 0 | \Psi_0\ra = 1/\sqrt{\mu}$ and $ \la 1 | \Psi_0\ra = 0 $. The latter relation implies that $ \la n | \Psi_0\ra = 0 $ whenever $n$ is odd, which is expected because the wave function associated with $|\Psi_0\ra$ is even under spatial parity $q\to-q$, while those associated with $|n\ra$ have parity $(-1)^n$. The solution of the recurrence relation Eq.~(\ref{recr}) is easily found in terms of double factorials as
\eq{\la n |\Psi_0\ra = \frac{(-\lambda)^{n/2}}{\sqrt{\mu}} \sqrt{\frac{(n-1)!!}{n!!}}.}
By using the property $\Gamma(k+1/2) = (2k-1)!!\sqrt{\pi}/2^k$ and $(2k)!!=2^k \Gamma(k+1)$ (see, e.g., 5.4.2 in Ref.~\onlinecite{HMF}), 
the solution for even $n$ is
\eq[ampl]{\la n | \Psi_0\ra = (-\lambda)^{n/2} \[ \frac{\Gamma\(\frac{n+1}{2}\)}{\mu\sqrt{\pi}\Gamma\(\frac{n}{2}+1\)}\]^{1/2},}
while $\la n | \Psi_0\ra = 0$ for all odd $n$. Substituting into Eq.~(\ref{a0}) we have
\eq[qqpf0FF]{G(t)=\frac{ {\rm e}^{i\o_0 t/2}}{\sqrt{\pi}\mu} \sum_{n=0}^\infty \frac{\Gamma\(n+\frac{1}{2}\)}{n!} {\rm e}^{-\(2n+\frac{1}{2}\) i \o t} \lambda^{2n}  .}
The sum of this series can be calculated by using the integral representation of $\Gamma$ and the result is 
\eq[qqpf0]{G(t) = \frac{ {\rm e}^{i(\o_0-\o)t/2}}{\mu \sqrt{1-\lambda^2 {\rm e}^{-2i\o t}}} = {\rm e}^{i(\o_0-\o)t/2} \sqrt{\frac{1-\lambda^2}{1-\lambda^2 {\rm e}^{-2i\o t}}} .}

We now repeat the derivation of Eq.~(\ref{qqpf0}) by using coherent states and then we extend it to the thermal case. The first observation is that the initial state for the post-quench Hamiltonian is a so-called \emph{squeezed vacuum}, defined as a state of the form $S(\xi)|0\ra$ where $S(\xi)\equiv\exp\left[\tfrac{1}{2}\left(\xi^* a^2-\xi a^{\dagger 2}\right)\right]$ is the unitary \emph{squeezing operator} and $|0\ra$ is the vacuum state after the quench. This is a consequence of the Bogoliubov relations Eq.~(\ref{Bog0}). In fact the squeezing operator $S=S(r {\rm e}^{i\phi})$ with $\cosh r = \mu$ and $ {\rm e}^{i\phi}\sinh r = \nu$ performs exactly the Bogoliubov rotation according to $a_0=S a S^\dg$, from which follows that the state $S|0\ra$ is annihilated by $a_0$ since $a_0 (S|0\ra) = (S a S^\dg) (S|0\ra) = S a |0\ra = 0$. Therefore $|\Psi_0\ra = S|0\ra$ or, equivalently,
\eq[in_st0]{|\Psi_0\ra =(1-\lambda^2)^{1/4} \exp\(-\frac{\lambda}{2}  a^{\dagger 2}\)|0\ra ,}
as it can be shown for example by normal-ordering the operator $S$ (\cite{squeezed}). One can verify directly that $a_0$ given by Eq.~(\ref{Bog0}) annihilates the state in Eq.~(\ref{in_st0}) using the identity $[a,f(a^\dagger)]=f'(a^\dagger)$.

The characteristic function $G(t)$ can now be calculated by using known results from the theory of coherent and squeezed coherent states. Introducing in $\la\Psi_0| {\rm e}^{-iHt}|\Psi_0\ra$ of Eq.~(\ref{0}) a resolution of the identity in the basis of coherent states of the post-quench Hamiltonian, we get
\eq[1]{G(t) = {\rm e}^{i\o_0 t/2} \int \frac{{\rm d}^2 z}{\pi} \la\Psi_0| {\rm e}^{-iHt}|z\ra \la z|\Psi_0\ra ,}
where $|z\ra$ indicates a coherent state ($a|z\ra=z|z\ra$ with complex $z$) and ${\rm d}^2 z \equiv {\rm d} \text{Re} z \; {\rm d} \text{Im} z$. The action of the evolution operator on the coherent states is simply
\eq[a1]{ {\rm e}^{-iHt}|z\ra = {\rm e}^{-i\o t/2}|z {\rm e}^{-i\o t}\ra ,}
as one can easily verify by expanding the coherent state $|z\ra$ on the basis of the eigenstates of $H$. Accordingly, from Eq.~(\ref{1}) one finds
\eq[2]{G(t) = {\rm e}^{i(\o_0-\o)t/2} \int \frac{\rmd^2 z}{\pi} \la\Psi_0|z {\rm e}^{-i\o t}\ra \la z|\Psi_0\ra .}
The overlap between a coherent state $|\alpha\ra = D(\alpha)|0\ra$ and a squeezed coherent state $|\beta;\xi\ra = S(\xi)D(\beta)|0\ra$, where $D(\alpha)=\exp(\alpha a^\dg - \alpha^* a)$ is the unitary \emph{displacement operator}, is given by\cite{VW94}
\eq[a3b]{\la\alpha|\beta;\xi\ra = \frac{1}{\sqrt{\mu}} {\rm e}^{ -\frac{1}{2}(|\alpha|^2+|\beta|^2)-\frac{1}{2\mu}(\nu\alpha^{*2}-\nu^*\beta^2-2\alpha^*\beta)},}
where, as before, $\mu=\cosh\rho$ and $\nu= {\rm e}^{i\theta}\sinh\rho$. A squeezed vacuum is a squeezed coherent state with $\beta=0$ and so in our case, according to the above relations, we have
\eq[a2b]{\la z|\Psi_0\ra = \frac{1}{\sqrt{\mu}} \exp\[-\frac{1}{2}(|z|^2 + \lambda z^{*2})\]. }
Alternatively, the last expression can be derived directly from Eq.~(\ref{in_st0}) and the definition of a coherent state $a|z\ra=z|z\ra$ along with the standard relation $ \la z|0\ra = {\rm e}^{-\frac12 |z|^2} $ determined by their normalization.

Substituting Eq.~(\ref{a2b}) in Eq.~(\ref{2}) we find
\eq{G(t) = \frac{ {\rm e}^{i(\o_0-\o)t/2}}{\pi\mu} \int {\rm d}^2 z \; \exp\[-|z|^2 -\frac{\lambda}{2}(z^2 {\rm e}^{-2i\o t} + z^{*2})\],}
which is simply a double  integral that can be evaluated straightforwardly, giving the final result
(\ref{qqpf0}).

Note that the calculation presented above for $\la \Psi_0| {\rm e}^{-iHt} |\Psi_0\ra$ can also be done in imaginary time, i.e., for $\langle \Psi_0| {\rm e}^{- H R}| \Psi_0\rangle$ which represents the partition function of the corresponding classical model in a film of width $R$ with boundary states both equal to $|\Psi_0\ra$, and provides the analytical continuation of the above result for $t=-iR$. The only difference is that, due to the standard normalization of coherent states, we now have
\eq[a2]{ {\rm e}^{-HR}|z\ra = {\rm e}^{-\tfrac{1}{2}\o R - \tfrac{1}{2} |z|^2 (1- {\rm e}^{-2 \o R})} \; |z {\rm e}^{-\o R}\ra }
instead of Eq.~(\ref{a1}).

Apart from the overall phase factor $ {\rm e}^{i(\o_0-\o)t/2}$ which corresponds to the difference between the ground-state energies of the pre- and post-quench Hamiltonian, the rest of Eq.~(\ref{qqpf0}) is a periodic function of $t$ with period $2\omega$ and as such it can be expanded as a Fourier series, from which we deduce that the probability distribution Eq.~(\ref{0a}) is given by an equidistant sequence of $\delta$-peaks at positions $W_n=(\o-\o_0)/2+2n\o$, $n=0,1,2,3,...$ (Fig.~\ref{sho}). More precisely, by expanding Eq.~(\ref{qqpf0}) as a Fourier series in $t$ (or Taylor expanding in powers of $\lambda^2 {\rm e}^{-2i\o t}$) we recover Eq.~(\ref{qqpf0FF}).

We now consider a quantum quench from $\o_0$ to $\o$ but starting from a thermal initial state with inverse temperature $\beta$, instead of the ground state. In this case the calculation in the Fock basis is significantly more difficult than before since knowledge of all transition amplitudes between eigenstates of the pre-quench Hamiltonian with those of the post-quench one is required\cite{Lutz1008}. The coherent state method instead remains relatively simple.

%
%%%%%%%%%%%%%%%%%%%%%%%%%%%%
\begin{figure}[ht]
\centering
\includegraphics[width=.95\columnwidth]{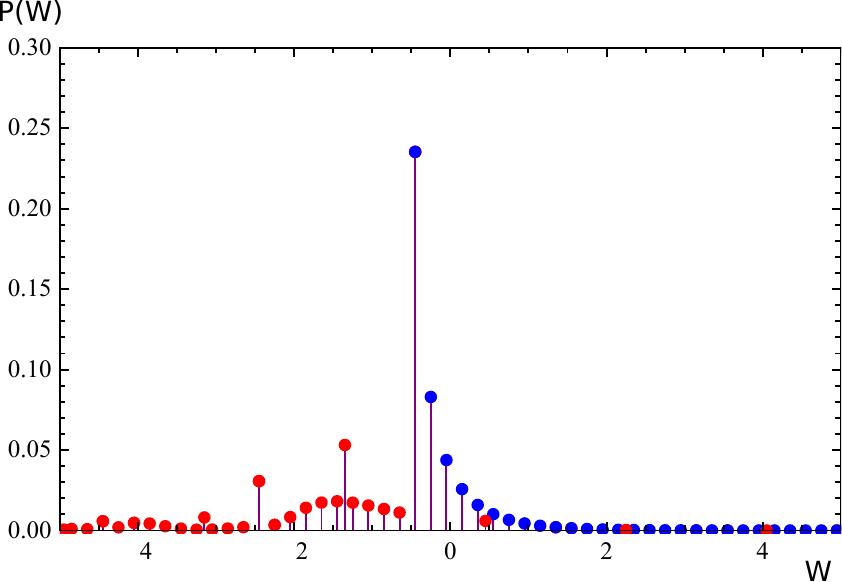}
\caption{\label{sho_th} \small Probability distribution $P(W)$ for a thermal quantum quench in a single harmonic oscillator with $\o=0.1$, $\o_0=1$, and $\beta=0.5$. This choice of $\omega$ and $\omega_0$ is the same as the one of the quench reported is Fig.~\ref{sho}, which the present one has to be compared with. The red dots indicate peaks that are of purely thermal origin.}
\end{figure}
%%%%%%%%%%%%%%%%%%%%%%%%%%%%
%
The characteristic function $G(t)$ is now given by
\eq{ G(t) = \frac{\text{Tr}\{ {\rm e}^{-\beta H_0} { {\rm e}^{iH_0 t}} {\rm e}^{-iHt} \}}{\text{Tr}\{ {\rm e}^{-\beta H_0} \}}.}
By introducing a resolution of the identity in the coherent state basis of the post-quench Hamiltonian, one finds
\eq{\text{Tr}\{ {\rm e}^{(-\beta+it) H_0} {\rm e}^{-iHt} \} = \int \frac{\rmd^2 z}{\pi} \la z| {\rm e}^{(-\beta {+it}) H_0}{\rm e}^{-iHt} |z\ra.}
As in the ordinary case, the action of the evolution operator $ {\rm e}^{-iHt}$ on $|z\ra$ is simply given by Eq.~(\ref{a1}), but in order to proceed further we need to introduce once more a resolution of the identity, this time in the coherent state basis of the pre-quench Hamiltonian
\eqa{& \text{Tr}\{ {\rm e}^{(-\beta {+it})  H_0} {\rm e}^{-iHt} \} = \nn \\
& = \iint \frac{\rmd^2 z}{\pi} \frac{\rmd^2 \zeta}{\pi} {\la z| {\rm e}^{(-\beta {+it})  H_0} |\zeta\ra}{\la \zeta| {\rm e}^{-iHt} |z\ra},}
where the states $|\zeta\ra$ are coherent states in the pre-quench basis (i.e., $a_0|\zeta\ra = \zeta|\zeta\ra$) and squeezed coherent states in the post-quench basis. The action of the operator $ {\rm e}^{(-\beta {+it})  H_0}$ on $|\zeta\ra$ is now also simply given by Eqs.~(\ref{a1}) and (\ref{a2}) and yields
\eqa{& \text{Tr}\{ {\rm e}^{(-\beta {+it})  H_0} {\rm e}^{-iHt} \} = {\rm e}^{-\beta \o_0/2} {\rm e}^{i({\o_0}-\o) t/2} \times \nn \\
& \times \iint \frac{\rmd^2 z}{\pi} \frac{\rmd^2 \zeta}{\pi} {\rm e}^{-\tfrac{1}{2} |\zeta|^2 (1- {\rm e}^{-2 \beta \o_0})} \la z|\zeta {\rm e}^{(-\beta {+it}) \o_0} \ra \la \zeta|z {\rm e}^{-i\o t}\ra,}
which is a quadruple  integral involving overlap amplitudes between coherent and squeezed coherent states given by Eq.~(\ref{a3b}). After some algebra and taking into account the normalization factor $1/\text{Tr}\{ {\rm e}^{-\beta H_0}\}$, we finally find
\begin{widetext}
\eq[tqpf0]{G_{\rm th}(t) =  \frac{ {\rm e}^{i({\o_0}-\o) t/2}(1- {\rm e}^{-\beta \o_0}) \sqrt{1-\lambda^2}}{\sqrt{(1 - \lambda^2 {\rm e}^{-2i\o t}) - 2 (1-\lambda^2) {\rm e}^{-\beta \o_0 + i(\o_0-\o)t} + {\rm e}^{-2 \beta\o_0} ( {\rm e}^{2i(\o_0-\o)t} - \lambda^2 {\rm e}^{2 i \o_0 t} )}}.}
\end{widetext}
One can verify that this last expression agrees with the more general one of Ref.~\onlinecite{Lutz1008}, Eq.~(17). Note that in the limit $\beta \to \infty$ this expression renders Eq.~(\ref{qqpf0}), as expected. In contrast to the ordinary quench case in Eq.~(\ref{qqpf0}), the Fourier series of Eq.~(\ref{tqpf0}) consists of double frequency oscillations since both $ {\rm e}^{i\o_0 t}$ and $ {\rm e}^{i\o t}$ are involved. In order to highlight the emergence of novel peaks of thermal nature, we report in  Fig.~\ref{sho_th} the probability distribution function $P(W)$ obtained from Eq.~\reff{tqpf0} for the single harmonic oscillator quenched from an initial state with $\omega_0=1$ and $\beta=0.5$ to a final state with $\omega = 0.1$, i.e., for the same values of parameters $\omega_0$ and $\omega$ as those of the ordinary quantum quench reported in Fig.~\ref{sho}. Each vertical line indicate the occurrence of a $\delta$-function at the corresponding frequency, with the (integrated) amplitude reported in the vertical axis. In particular, the peaks highlighted in red are of purely thermal origin and are therefore absent in Fig.~\ref{sho}, together with all the peaks occurring for energies below the lower threshold $(\omega-\omega_0)/2$ which characterize the ordinary quantum quench. In addition, compared to the latter case (see Eq.~\reff{qqpf0}), 
$ {\rm e}^{i\o t}$ in Eq.~\reff{tqpf0} appears not only in even powers as before (reflecting the fact that the initial state $|\Psi_0\ra$ contained excitations of the post-quench Hamiltonian only in pairs), but also in odd power terms (meaning that the thermal initial state contains all levels of excitations), with the latter decaying for small temperatures ($\beta\to\infty$).

%\bibliography{Nonequilibrium}
%

\end{document}